\documentclass{amsart}

\usepackage[cp1250]{inputenc}

\usepackage{amsmath}
\usepackage{amsthm}
\usepackage{amssymb}

\usepackage{cite}

\usepackage{array}

\usepackage{bbm}

\usepackage{graphicx}

\usepackage{longtable}

\newcommand{\wektor}[1]{#1}
\newcommand{\linspan}[1]{\textnormal{span}\left(#1\right)}
\newcommand{\innerpr}[2]{\left<#1,#2\right>}
\newcommand{\proj}[2]{\left|#1\right>\left<#2\right|}
\newcommand{\kernel}[1]{\textnormal{Ker}#1}
\newcommand{\range}[1]{R\left(#1\right)}
\newcommand{\rank}[1]{r\left(#1\right)}
\newcommand{\SL}{\textnormal{SL}}
\newcommand{\PSL}{\textnormal{PSL}}
\newcommand{\PSLt}{\PSL\left(3,\mathbbm{C}\right)}
\newcommand{\SLt}{\SL\left(3,\mathbbm{C}\right)}
\newcommand{\PSLtt}{\PSLt\times\PSLt}
\newcommand{\SLtt}{\SLt\otimes\SLt}

\newtheorem{theorem}{Theorem}[section]
\newtheorem{lemma}[theorem]{Lemma}
\newtheorem{corollary}[theorem]{Corollary}
\newtheorem{remark}[theorem]{Remark}
\newtheorem{definition}[theorem]{Definition}
\newtheorem{example}[theorem]{Example}
\newtheorem{proposition}[theorem]{Proposition}

\author{Łukasz Skowronek}
\address{Instytut Fizyki im. Smoluchowskiego, Uniwersytet Jagielloński, Reymonta 4, 30-059 Kraków, Poland}
\curraddr{Stockholms universitet, Fysikum, 106 91 Stockholm, Sweden}
\email{lukasz.skowronek@uj.edu.pl}

\title[Bound entanglement with general unextendible product bases]{Three-by-three bound entanglement with general unextendible product bases}

\begin{document}
\begin{abstract}
We discuss the subject of Unextendible Product Bases with the orthogonality condition dropped and we prove that the lowest rank non-separable positive-partial-transpose states, i.e. states of rank $4$ in $3\times 3$ systems are always locally equivalent to a projection onto the orthogonal complement of a linear subspace spanned by an orthogonal Unextendible Product Basis. The product vectors in the kernels of the states belong to a non-zero measure subset of all general Unextendible Product Bases, nevertheless they can always be locally transformed to the orthogonal form. This fully confirms surprising numerical results recently reported by Leinaas et al. Parts of the paper rely heavily on the use of Bezout's Theorem from algebraic geometry.
\end{abstract}
\maketitle
\section{Introduction}
Unextendible Product Bases (UPBs for short) consisting of orthogonal vectors are well recognized by the quantum information community \cite{Bennett99,DiVicenzo04,Bravyi_2003,De_Rinaldis_2003,Cohen_2007,Duan_Xin_Ying_2007,ASHKLA11} and among mathematicians \cite{AL2001,Feng2006}, but similar product bases with the orthogonality condition dropped seem to have received very little attention \cite{Pittenger03,LS2010,LS2011} so far. This is partly explained by the fact that an orthogonal UPB can always be used to construct a PPT bound entangled state \cite{Bennett99}, while no simple prescription is known for general UPBs, cf. \cite{Pittenger03,HHMS2011}. Nevertheless, the recent results of \cite{LS2010}, if not plain curiosity, seem to call for a more systematic treatment of the subject of general UPBs. It has been observed in numerical searches \cite{LS2010} that non-separable positive-partial-transpose states (PPT states for short) of rank four in $3\times 3$ systems can be locally, i.e. by $\SLtt$ tranformations, brought to the form of an orthogonal projection onto the orthogonal complement of an orthogonal $3\times 3$ UPB \cite{DiVicenzo04}. The product vectors in the kernel of such PPT states belong to a non-zero measure subset of the set of all general $3\times 3$ UPBs with a minimal number of elements.  A valuable discussion of the subject has been included in the recent works \cite{HHMS2011,LS2011}. In the present paper, we provide an analytical proof of the main result of \cite{LS2010} and explain a number of other properties of extreme PPT states, obtained numerically and reported in \cite{LS2010n}. As a byproduct, we deliver a number of rather general techniques to tackle problems related to product vectors in the kernel of a PPT state. Facts from algebraic geometry, and specifically the Bezout's Theorem are of much use here. A recognized textbook on the subject is \cite{Harris}, and we are going to use it as a main reference.

The content of the paper is divided into two sections, each of them consisting of smaller pieces. The first section discusses the subject of general Unextendible Product Bases, without any constraints on dimensionality, neither on the number of subsystems. It also summarizes what is generally known for $2\times n$ systems as a consequence of the recent work \cite{ATL2010}, and partly also by the early findings of \cite{Bennett99}. The second part of the paper is devoted to the derivation of our core result, concerning PPT bound entangled states of minimal rank in $3\times 3$ systems. As we mentioned above, they are all locally transformed projections onto the orthogonal complement of a subspace spanned by an orthogonal UPB. On the way to prove this result, we give numerous insights into how PPT states relate to product vectors in their kernels. That there is a strong relation between this subject and general Unextendible Product Bases becomes clear in Section \ref{secgUPB}, where we prove that there always is a minimal gUPB in the kernel of a non-separable PPT state of rank $4$ in $3\times 3$ systems, and we make much use of the general results of the first part of the paper. A common trait in all the following is also an idea of local $\SLtt$ equivalence, explained with considerable care in Section \ref{secequivalence}. The equivalence is natural both for gUPBs and positive-partial-transpose states, as well as in algebraic geometry when the Segre variety is concerned. It should also be noticed that we make a number of comments, related to the results of \cite{LS2010n, ATL2010} and the question of atomicity of entanglement witnesses \cite{Ha98,Terhal2001}. They are somewhat out of the main focus in this paper, but in our opinion they may prove valuable for the reader.

Concerning the notation and language we use, only few things may need an explanation. Throughout the paper, $\left\{e_i\right\}_{i=1}^n$ denotes the canonical basis of $\mathbbm{C}^n$. That is, $e_i=\left[\begin{array}{ccccccc}0&\ldots&0&1&0&\ldots&0\end{array}\right]$ with $1$ at the $i$-th position and zeros elsewhere. We denote by $\kernel{\rho}$, $\range{\rho}$ and $\rank{\rho}$ the kernel, the range and the rank of a state $\rho$. The symbols $T_1$ and $T_2$ stand for the partial transpose with respect to the first and the second subsystem, respectively. For example, $\rho^{T_2}$ is a state $\rho$, partially transposed with respect to the second subsystem. We will be using the term ``edge state'' from \cite{LKHC2001}. A PPT edge state $\rho$ is by definition a $\rho$ which does not admit $\phi\otimes\psi\in\range{\rho}$ such that $\phi^{\ast}\otimes\psi\in\range{\rho^{T_1}}$. Equivalently, there is no $\phi\otimes\psi\in\range{\rho}$ such that $\phi\otimes\psi^{\ast}\in\range{\rho^{T_2}}$. We say that a state $\rho$, acting on $\mathbbm{C}^n\otimes\mathbbm{C}^m$, is supported on a $k\times l$ subspace when there exist a $k$-dimensional subspace $\mathcal{L}_1$ of $\mathbbm{C}^n$ and an $l$-dimensional subspace $\mathcal{L}_2$ of $\mathbbm{C}^m$ such that $\range{\rho}\subset\mathcal{L}_1\otimes\mathcal{L}_2$, while there exists no similar subspace  $\mathcal{L}'_1\otimes\mathcal{L}'_2$ with $\dim\mathcal{L}'_1<l$ or $\dim\mathcal{L}'_2<k$. We should also add that the notation $\Sigma_{n-1,m-1}$, in accordance with \cite{Harris}, is used for the Segre variety in $\mathbbm{P}^{nm-1}$, or less precisely, for the set of product vectors in $\mathbbm{C}^n\otimes\mathbbm{C}^m$. We do not distinguish much between the two.

\section{General Unextendible Product Bases}
\subsection{Basic results}
Let us start with a definition of a general UPB.
\begin{definition}\label{defnonortUPB} Take $n,m\in\mathbbm{N}$. By a {\bf general Unextendible Product Basis}, or a gUPB for short, we mean a set  $\left\{\wektor{\phi_i\otimes\psi_i}\right\}_{i=1}^k$ of product vectors in $\mathbbm{C}^n\otimes\mathbbm{C}^m$, $0<k<mn$, such that there is no product vector in $\linspan{\left\{\wektor{\phi_i\otimes\psi_i}\right\}_{i=1}^k}^{\bot}$, the orthogonal complement to the linear span of $\left\{\wektor{\phi_i\otimes\psi_i}\right\}_{i=1}^k$.
\end{definition}
In other words, a gUPB is a set of product vectors $\left\{\wektor{\phi_i\otimes\psi_i}\right\}_{i=1}^k$ such that there is no product vector orthogonal to all of them. Yet another way of phrasing the definition is that the orthogonal complement to a gUPB is a Completely Entangled Subspace \cite{Partha04}, or CES for short.

\begin{remark}The definition of a gUPB can be trivially extended to a multipartite setting.\end{remark}

We know that gUPBs do exist. Any UPB consisting of orthogonal vectors is an example (concrete UPBs can be found e.g. in  \cite{DiVicenzo04}). We also know that for some spaces, no UPB consisting of orthogonal vectors can exist. For example, it has been noticed as early as in \cite{Bennett99} that $2\times n$ systems do not admit an orthogonal UPB, and a more general discussion of existence questions for orthogonal UPBs has been included in \cite{AL2001}. In the following, we show that gUPBs are much more common than the usual UPBs, and give a characterization of gUPBs of minimal number of elements. In Section \ref{seclowdim} we discuss in more detail the $2\times n$ case, where much is known as a consequence of the recent work \cite{ATL2010}. 

First, let us answer a question about the minimum number of elements in a gUPB in $\mathbbm{C}^n\otimes\mathbbm{C}^m$. 

\begin{proposition}\label{propgUPBnonexistence}A set of vectors $\left\{\wektor{\phi_i\otimes\psi_i}\right\}_{i=1}^k\subset\mathbbm{C}^n\otimes\mathbbm{C}^m$ consisting of $k<m+n-1$ elements is not a generalized UPB.
\begin{proof} There exists a vector $\wektor{f}\in\mathbbm{C}^n$ orthogonal to all the vectors $\wektor{\phi_i}$ with $i=1,\ldots,n-1$. Moreover, there exists a $\wektor{g}\in\mathbbm{C}^m$ orthogonal to the vectors $\wektor{\psi_j}$ with $j=n,\ldots,k$ (because $k-n<m$). The vector $\wektor{f\otimes g}$ is orthogonal to all $\wektor{\phi_i\otimes\psi_i}$ for $i=1,\ldots, k$.
\end{proof}
\end{proposition}

\begin{proposition}\label{propgUPBsuffnecessary}
A set of vectors $\left\{\wektor{\phi_i\otimes\psi_i}\right\}_{i=1}^{m+n-1}\subset\mathbbm{C}^n\otimes\mathbbm{C}^m$ is a gUBP if and only if any $n$-tuple of vectors in $\left\{\wektor{\phi_i}\right\}_{i=1}^{m+n-1}$ consists of linearly independent vectors, as well as any $m$-tuple of vectors in $\left\{\wektor{\psi_j}\right\}_{j=1}^{m+n-1}$.
\begin{proof}
In order to prove necessity, assume that an $n$-tuple of vectors $\left\{\wektor{\phi_{i_l}}\right\}_{l=1}^n$ is linearly dependent. Therefore there exists a vector $\wektor{f}\in\mathbbm{C}^n$ orthogonal to all of them. Vectors of the form $\wektor{f\otimes g}\in\mathbbm{C}^n\otimes\mathbbm{C}^m$ with an arbitrary $\wektor{g}\in\mathbbm{C}^m$ are orthogonal to all the vectors $\left\{\wektor{\phi_{i_l}\otimes\psi_{i_l}}\right\}_{l=1}^n\subset\mathbbm{C}^n\otimes\mathbbm{C}^m$. Obviously, $\wektor{g}$ can be chosen in such a way that $\wektor{f\otimes g}$ is orthogonal to the remaining $m-1$ elements of $\mathbbm{C}^m$ (because $m-1<m=\dim\mathbbm{C}^m$). For a sufficiency proof, assume that $\wektor{f\otimes g}$ is orthogonal to $\wektor{\phi_i\otimes\psi_i}$ for $i=1,\ldots,m+n-1$. The vector $\wektor{f}$ can be orthogonal to at most $n-1$ of $\wektor{\phi_i}$'s, whereas $\wektor{g}$ cannot be orthogonal to more than $m-1$ $\wektor{\psi_i}$'s (remember the linear independence of $n$-tuples and $m$-tuples, respectively). This gives a maximum of $\left(n-1\right)+\left(m-1\right)$ vectors in $\left\{\wektor{\phi_i\otimes\psi_i}\right\}_{i=1}^{m+n-1}$ orthogonal to $\wektor{f\otimes g}$. Therefore $\wektor{f\otimes g}$ cannot be orthogonal to all the $\wektor{\phi_i\otimes\psi_i}$'s, the set $\left\{\wektor{\phi_i\otimes\psi_i}\right\}_{i=1}^{m+n-1}$ is a gUPB.
\end{proof}
\end{proposition}

As we already mentioned above, it is a trivial observation that the orthogonal complement to a gUPB is a Completely Entangled Subspace \cite{Partha04}. However, it is more interesting to notice that the minimum number $m+n-1$ of elements in a generalized UPB exactly matches a factor which appears in the formula $D=mn-\left(m+n-1\right)$ for the maximum dimension $D$ of a CES in $\mathbbm{C}^n\otimes\mathbbm{C}^m$ \cite{Wallach02,Partha04,Cubitt07}. This may seem a little bit puzzling at first, but can be easily explained. Consider a generic set of $m+n-2$ vectors $\left\{\phi_i\otimes\psi_i\right\}_{i=1}^{m+n-2}\subset\mathbbm{C}^n\otimes\mathbbm{C}^m$. It will have the property that any $n$-tuple in $\left\{\phi_i\right\}_{i=1}^{m+n-2}$ and any $m$-tuple in $\left\{\psi_j\right\}_{j=1}^{m+n-2}$ are linearly independent. Therefore if there is a vector $\phi\otimes\psi$ orthogonal to all the $\phi_i\otimes\psi_i$'s, we must have $\innerpr{\phi}{\phi_i}=0$ for exactly $n-1$ vectors $\phi_i$ and $\innerpr{\psi}{\psi_j}=0$ for exactly to $m-1$ vectors $\psi_j$. More precisely, there must exist $\mathcal{N},\mathcal{M}\subset\left\{1,2,\ldots,m+n-2\right\}$ such that $\#\mathcal{N}=n-1$, $\#\mathcal{M}=m-1$, $\mathcal{N}\cup\mathcal{M}=\left\{1,2,\ldots,m+n-2\right\}$ and $\innerpr{\phi}{\phi_i}=0\,\forall_{i\in\mathcal{N}}$, as well as $\innerpr{\psi}{\psi_j}=0\,\forall_{j\in\mathcal{M}}$. Moreover, for any choice of an $\left(n-1\right)$-tuple $\mathcal{N}$ of vectors $\phi_i$ and a complementary $\left(m-1\right)$-tuple $\mathcal{M}$ of vectors $\psi_j$, there exists exactly one product vector $\phi\otimes\psi$ with $\innerpr{\phi}{\phi_i}=0\,\forall_{i\in\mathcal{N}}$ and $\innerpr{\psi}{\psi_j}=0\,\forall_{j\in\mathcal{M}}$. Altogether, we have $\binom{m+n-2}{m-1}$ suitable pairs $\left(\mathcal{N},\mathcal{M}\right)$ of $\left(n-1\right)$ and $\left(m-1\right)$-tuples, which gives us exactly $\binom{m+n-2}{m-1}$ product vectors orthogonal to all the $\phi_i\otimes\psi_i$'s. To no surprise, this number equals the degree of the respective Segre variety $\Sigma_{n-1,m-1}$, cf. \cite[Lecture 18]{Harris}.  A key fact for us is that the number of product vectors is finite. If we assume that $\linspan{\left\{\phi_i\otimes\psi_i\right\}_{i=1}^{m+n-2}}^{\bot}$ and $\Sigma_{n-1,m-1}$ intersect generically enough \cite[Example 18.5]{Harris}, we are lead to a conclusion that the projective dimension $mn-1-\left(m+n-2\right)$ of the orthogonal complement to the $\phi_i\otimes\psi_i$'s must be bigger exactly by one than the projective dimension $D-1$ of a maximal CES. Thus we get $D=mn-\left(m+n-2\right)-1=mn-\left(m+n-1\right)$, as expected. Note that this is only a rough explanation and not a rigorous proof. We also use it as a good opportunity to introduce and explain the formula for the degree of the Segre variety, which we are going implicitly refer to later.

It is intuitively clear that the $n$-tuple ($m$-tuple) linear independence property of Proposition \ref{propgUPBsuffnecessary} is generic for random finite sets of vectors in $\mathbbm{C}^n$ ($\mathbbm{C}^m$, resp.) and therefore minimal gUPBs should exist for any $m$, $n$ (one can check it numerically, for example). However, it does not seem entirely trivial to prove this fact in a mathematically rigorous way. We do it in the following proposition.

\begin{proposition}\label{propgUPBexistence}For arbitrary $n,m\in\mathbbm{N}$, there exists a general UPB in $\mathbbm{C}^n\otimes\mathbbm{C}^m$ consisting of $m+n-1$ vectors. 
\begin{proof}
According to Proposition \ref{propgUPBsuffnecessary}, it is sufficient to prove that there exists a set of $m+n-1$ vectors $\wektor{\phi_i}\in\mathbbm{C}^n$ s.t. any $n$-tuple is linearly independent, and similarly a set of $m+n-1$ vectors $\wektor{\psi_i}\in\mathbbm{C}^m$ with linearly independent $m$-tuples. We prove a more general statement in the following lemma.
\begin{lemma}\label{lemma1}
For any $n\in\mathbbm{N}$, there exists an infinite sequence $\left\{\wektor{\phi_i}\right\}_{i=1}^{+\infty}$ of vectors in $\mathbbm{C}^n$ such that any $n$-tuple of them is linearly independent.
\begin{proof}
Consider the family of vectors
\begin{equation}\label{eqvecVandermonde}
\wektor{\phi_i}=\left[\begin{array}{ccccc}1&\alpha_i&\alpha_i^2&\ldots&\alpha_i^{n-1}\end{array}\right],
\end{equation}
where $\left\{\alpha_i\right\}_{i=1}^{+\infty}$ is a sequence of numbers (real or complex) such that $\alpha_i\neq\alpha_j$ for $i\neq j$. Let us choose some $\left\{i_1,i_2,\ldots,i_n\right\}\in\mathbbm{N}$ s.t. $i_j\neq i_k$ for $j\neq k$. The matrix
\begin{equation}\label{eqVandermonde}
\left[\begin{array}{ccccc}
1 & \alpha_{i_1} & \alpha_{i_1}^2 & \hdots & \alpha_{i_1}^{n-1}\\
1 & \alpha_{i_2} & \alpha_{i_2}^2 & \hdots & \alpha_{i_2}^{n-1}\\
\vdots & \vdots & \vdots & \ddots & \vdots\\
1 & \alpha_{i_n} & \alpha_{i_n}^2 & \hdots & \alpha_{i_n}^{n-1}\\
\end{array}
\right]
\end{equation}
is a Vandermonde matrix \eqref{eqVandermonde}, and it is well known that its determinant equals 
\begin{equation}
\prod_{0\leqslant j<k\leqslant n}\left(\alpha_{i_j}-\alpha_{i_k}\right).
\end{equation}
Therefore, the matrix is nonsingular for any choice of the $n$-tuple $\left\{i_1,i_2,\ldots,i_n\right\}\in\mathbbm{N}$. Equivalently, any $n$-tuple of vectors $\wektor{\phi_i}$ from equation \eqref{eqvecVandermonde} is linearly independent.
\end{proof} 
\end{lemma} 
From Lemma \ref{lemma1}, we almost immediately arrive at the proposition. Choose $m+n-1$ vectors $\wektor{\phi_i}$ as in equation \eqref{eqvecVandermonde}. In a similar fashion, take
\begin{equation}\label{eqvecVandermonde2}
\wektor{\psi_i}=\left[\begin{array}{ccccc}1&\beta_i&\beta_i^2&\ldots&\beta_i^{m-1}\end{array}\right],
\end{equation}
where again $\beta_i\neq\beta_j$ for $i\neq j$. By Lemma \ref{lemma1} and Proposition \ref{propgUPBsuffnecessary}, the set $\left\{\wektor{\phi_i\otimes\psi_i}\right\}_{i=1}^{m+n-1}$ constitutes a gUBP. 
\end{proof}
\end{proposition}

It is natural to ask for a generalization of Proposition \ref{propgUPBsuffnecessary} for sets of product vectors consisting of more than $m+n-1$ elements. We have the following

\begin{proposition}\label{progUPBnecsuffnonminimal}
A set of vectors $\left\{\wektor{\phi_i\otimes\psi_i}\right\}_{i=1}^N\subset\mathbbm{C}^n\otimes\mathbbm{C}^m$ with $N\geqslant m+n-1$ is a gUBP if and only if for any $\mathcal{N},\mathcal{M}\subset\mathbbm{N}$ such that $\mathcal{N}\cap\mathcal{M}=\emptyset$ and $\mathcal{N}\cup\mathcal{M}=\left\{1,2,\ldots,N\right\}$, at least one of the sets of vectors $\left\{\phi_i\right\}_{i\in\mathcal{N}}$ and $\left\{\psi_j\right\}_{j\in\mathcal{M}}$ spans the entire corresponding vector space (\/$\mathbbm{C}^n$ or $\mathbbm{C}^m$, resp.).
\begin{proof}Let us first prove necessity. Assume that the vectors $\left\{\phi_i\otimes\psi_i\right\}_{i=1}^N$ constitute a gUPB and choose some $\mathcal{N},\mathcal{M}\subset\mathbbm{N}$ as in the statement of the proposition. If neither of the sets $\left\{\phi_i\right\}_{i\in\mathcal{N}}$ and $\left\{\psi_j\right\}_{j\in\mathcal{M}}$ spans the respective vector space, there exist $\wektor{f}\in\mathbbm{C}^n$ and $\wektor{g}\in\mathbbm{C}^m$ such that $\innerpr{f}{\phi_i}=0$ and $\innerpr{g}{\psi_j}=0$ for all $i\in\mathcal{N}$ and $j\in\mathcal{M}$. Because of the condition $\mathcal{N}\cup\mathcal{M}=\left\{1,2,\ldots,N\right\}$, we clearly have $\innerpr{f\otimes g}{\phi_i\otimes\psi_i}=0$ for $i=1,2,\ldots,N$. This clearly contradicts the fact that the vectors $\phi_i\otimes\psi_i$ constitute a gUPB. In order to show sufficiency, assume that $f\otimes g\in\mathbbm{C}^n\otimes\mathbbm{C}^m$ is such that $\innerpr{f\otimes g}{\phi_i\otimes\psi_i}=0$ for all $i=1,2,\ldots,N$. Define $\mathcal{N}_f:=\left\{i\,\vline\innerpr{f}{\phi_i}=0\right\}$ and $\mathcal{M}_g:=\left\{j\,\vline\innerpr{g}{\psi_j}=0\right\}$. Clearly, we must have $\mathcal{N}_f\cup\mathcal{M}_g=\left\{1,2,\ldots,N\right\}$. Thus it is possible to choose $\mathcal{N}\subset\mathcal{N}_f$ and $\mathcal{M}\subset\mathcal{M}_g$ such that $\mathcal{N}\cup\mathcal{M}=\left\{1,2,\ldots,N\right\}$ and $\mathcal{N}\cap\mathcal{M}=\emptyset$. By the very definition of $\mathcal{N}_f$ and $\mathcal{M}_g$, we have $\innerpr{f}{\phi_i}=0$ for all $i\in\mathcal{N}$ and $\innerpr{g}{\psi_j}$ for  $j\in\mathcal{M}$. But according to the assumptions of the theorem, this is only possible if $f$ and $g$ are equal to zero. Thus $\left\{\phi_i\otimes\psi_i\right\}_{i=1}^N$ is a gUPB.
\end{proof}
\end{proposition}

\begin{remark}For $N=m+n-1$, Proposition \ref{progUPBnecsuffnonminimal} is the same as Proposition \ref{propgUPBsuffnecessary}, which should be expected.
\end{remark}

\begin{remark}For a given set of vectors $\left\{\phi_i\otimes\psi_i\right\}_{i=1}^N$, $N\geqslant m+n-1$, the condition of Proposition \ref{progUPBnecsuffnonminimal} can easily be checked, e.g. using a computer algebra program.
\end{remark}

Certain characterisations of gUPBs were earlier obtained in \cite{Pittenger03}, but the above results were, rather surprisingly, never explicitly given. They can also easily be generalized to a multipartite setting.

\begin{proposition}\label{progUPBnecsuffmulti}
A set of vectors $\left\{\wektor{\phi^1_i\otimes\phi^2_i\otimes\ldots\otimes\phi^l_i}\right\}_{i=1}^N\subset\mathbbm{C}^{n_1}\otimes\mathbbm{C}^{n_2}\otimes\ldots\mathbbm\otimes{C}^{n_l}$ with $N\geqslant\sum_{i=1}^ln_i+l-1$ is a gUBP if and only if for any $\mathcal{N}_1,\ldots,\mathcal{N}_l\subset\mathbbm{N}$ such that $\mathcal{N}_i\cap\mathcal{N}_j=\emptyset$ for all $i\neq j$ and $\bigcup_{i=1}^l\mathcal{N}_i=\left\{1,2,\ldots,N\right\}$, at least one of the sets of vectors $\left\{\phi^i_j\right\}_{j\in\mathcal{N}_i}$, $i=1,2,\ldots,l$, spans the entire corresponding vector space $\mathbbm{C}^{n_i}$.
\begin{proof}
Follows the same lines as the proof of Proposition \ref{progUPBnecsuffnonminimal} and will be omitted here.
\end{proof}
\end{proposition}

Note that a gUPB consisting of more than $m+n-1$ elements does not have to contain a minimal one. This can be seen in the following example.

\begin{example}\label{ex2x3}Choose the following five vectors in $\mathbbm{C}^2\otimes\mathbbm{C}^3$,
\begin{equation}\label{vectors2x3}
\left\{e_1\otimes e_1,e_2\otimes e_2,\left(e_1+e_2\right)\otimes e_3,e_2\otimes\left(e_1+e_2+e_3\right),e_1\otimes\left(e_1+2e_2-e_3\right)\right\}.
\end{equation}
\end{example}
It is a simple exercise to check that the set \eqref{vectors2x3} satisfies the assumptions of Proposition \ref{progUPBnecsuffnonminimal}, but neither of its four-element subsets is a minimal gUPB. However, it is still an open question whether there exists a minimal gUBP in the linear span of the vectors \eqref{vectors2x3}. In fact it can easily be checked that the following four vectors belong to the linear span of \eqref{vectors2x3} and they constitute a minimal gUPB.
\begin{eqnarray}\label{eqminimalgUPB2x31}
\left(e_1-e_2\right)\otimes\left(e_1-2e_2+e_3\right)&\left(e_1+e_2\right)\otimes\left(e_1-2e_2+2e_3\right)\\
 e_1\otimes\left(e_1-6e_2+3e_3\right)&\left(e_1+2e_2\right)\otimes\left(e_1-4e_2\right)\label{eqminimalgUPB2x32}
\end{eqnarray}
Thus, by giving Example \ref{ex2x3} we have not proved whether there exist a linear subspace complementary to a Completely Entangled Subspace \cite{Partha04} and  without a minimal gUPB in it or not. This seems to be an interesting question on its own.

\subsection{The simplest case}\label{seclowdim}
In the following, we are going to shortly discuss the $2\times n$ case, where much about generalized Unextendible Product Bases follows from the recent work \cite{ATL2010}. Let us however begin with something which has been known for much longer time \cite{Bennett99}.

Orthogonal UPBs in $2\times n$ systems do not exist. In \cite{Bennett99}, it followed from a local measurement protocol, but a direct proof can also be given, and it seems rather instructive to provide it here.
\begin{proposition}\label{propnoUPB2xn}Let $n\in\mathbbm{N}$ be an arbitrary positive integer. A set of mutually orthogonal product vectors $\left\{\wektor{\phi_i\otimes\psi_i}\right\}_{i=1,\ldots,N}\subset\mathbbm{C}^2\otimes\mathbbm{C}^n$ either spans the whole $\mathbbm{C}^2\otimes\mathbbm{C}^n$, or it admits a product vector $\wektor{f\otimes g}\in\mathbbm{C}^2\otimes\mathbbm{C}^n$, orthogonal to all the $\wektor{\phi_i\otimes\psi_i}$'s.
\begin{proof}For an inductive proof, let us first consider the case $n=1$. By Proposition \ref{propgUPBsuffnecessary}, a set of product vectors $\left\{\wektor{\phi_i\otimes e_1}\right\}_{i=1,\ldots,N}$ in $\mathbbm{C}^2\otimes\mathbbm{C}^1$ is a gUPB if and only if the vectors $\wektor{\phi_i}$ span $\mathbbm{C}^2$. Equivalently, $\wektor{\phi_i\otimes e_1}$'s span $\mathbbm{C}^2\otimes\mathbbm{C}^1$.

Now assume that the proposition holds for all $n=1,2,\ldots,p-1$, $p>1$ and consider a gUPB $\left\{\wektor{\phi_i\otimes\psi_i}\right\}_{i=1,\ldots,N}$ in $\mathbbm{C}^2\otimes\mathbbm{C}^p$, consisting of mutually orthogonal vectors. Define $\mathcal{I}:=\left\{i\,\vline\,\phi_i\sim\phi_1\right\}$, $\mathcal{J}:=\left\{j\,\vline\,\phi_j\bot\phi_1\right\}$, and $\mathcal{K}$ as the complement of $\mathcal{I}\cup\mathcal{J}$ in $\left\{1,2,\ldots,N\right\}$. By choosing $\mathcal{N}=\mathcal{I}$ in Proposition \ref{progUPBnecsuffnonminimal}, we see that the vectors $\left\{\psi_j\right\}_{j\in\mathcal{J}\cup\mathcal{K}}$ span $\mathbbm{C}^p$. By taking $\mathcal{N}=\mathcal{J}$, we arrive at a similar conclusion for  $\left\{\psi_i\right\}_{i\in\mathcal{I}\cup\mathcal{K}}$. Moreover, the vectors within the group $\left\{\psi_i\right\}_{i\in\mathcal{I}}$ have to be orthogonal, and similarly for $\left\{\psi_j\right\}_{j\in\mathcal{J}}$. Both sets $\left\{\psi_i\right\}_{i\in\mathcal{I}}$ and $\left\{\psi_j\right\}_{j\in\mathcal{J}}$ also need to be orthogonal to $\left\{\psi_k\right\}_{k\in\mathcal{K}}$. Taking all these facts about $\left\{\psi_i\right\}_{i\in\mathcal{I}}$ and $\left\{\psi_j\right\}_{j\in\mathcal{J}}$ together, we have $\linspan{\left\{\psi_i\right\}_{i\in\mathcal{I}}}=\linspan{\left\{\psi_j\right\}_{j\in\mathcal{J}}}={\linspan{\left\{\psi_k\right\}_{k\in\mathcal{K}}}}^{\bot}$, as well as $\#\mathcal{I}=\#\mathcal{J}$. In the case $\#\mathcal{I}=\#\mathcal{J}=p$, we have $N=2p$, $\mathcal{K}=\emptyset$, and the vectors $\left\{\wektor{\phi_i\otimes\psi_i}\right\}_{i=1,\ldots,N}$ span $\mathbbm{C}^2\otimes\mathbbm{C}^p$. If $\mathcal{K}\neq\emptyset$, $\left\{\wektor{\phi_k\otimes\psi_k}\right\}_{k\in\mathcal{K}}$ is a gUPB in $\mathbbm{C}^2\otimes\linspan{\left\{\psi_k\right\}_{k\in\mathcal{K}}}$, which is equivalent to $\mathbbm{C}^2\otimes\mathbbm{C}^n$ for some $n<p$. By our inductive assumption, $\left\{\wektor{\phi_k\otimes\psi_k}\right\}_{k\in\mathcal{K}}$ span $\mathbbm{C}^2\otimes\linspan{\left\{\psi_k\right\}_{k\in\mathcal{K}}}$. On the other hand, it is easy to see that the remaining product vectors $\left\{\wektor{\phi_i\otimes\psi_i}\right\}_{i\in\mathcal{I}\cup\mathcal{J}}$ span $\mathbbm{C}^2\otimes\left(\linspan{\left\{\psi_k\right\}_{k\in\mathcal{K}}}\right)^{\bot}$. Altogether, the vectors $\left\{\wektor{\phi_i\otimes\psi_i}\right\}_{i=1,2,\ldots,N}=\left\{\wektor{\phi_i\otimes\psi_i}\right\}_{i\in\left(\mathcal{I}\cup\mathcal{J}\right)\cup\mathcal{K}}$ span $\mathbbm{C}^2\otimes\mathbbm{C}^p$, which was our assertion.
\end{proof}
\end{proposition}

Let us now state the following lemma, which is simply an amalgam of Lemmas~1 and 2 in a paper by other authors \cite{ATL2010}.
\begin{lemma}\label{lemmaAugusiak}
Let $V$ be a Completely Entangled Subspace in $\mathbbm{C}^2\otimes\mathbbm{C}^n$. The product vectors $\phi\otimes\psi$ orthogonal to $V$ span $V^{\bot}$, while their partial conjugates $\phi^{\ast}\otimes\psi$ span  $\mathbbm{C}^2\otimes\mathbbm{C}^n$.
\end{lemma}
We immediately get the following
\begin{corollary}\label{corspanningces}
Let $V\subset\mathbbm{C}^2\otimes\mathbbm{C}^n$ be a CES of dimension $n-k$, $k\geqslant 1$. The orthogonal complement $V^{\bot}$ admits a general Unextendible Product Basis, consisting of $2n-(n-k)=n+k$ elements.
\begin{proof}
By Lemma \ref{lemmaAugusiak} we know that it is possible to find a basis $\left\{\phi_i\otimes\psi_i\right\}_{i=1}^{n+k}$ of $V^{\bot}$. Since $V$ is a CES, $\left\{\phi_i\otimes\psi_i\right\}_{i=1}^{n+k}$ must be a gUPB.
\end{proof}
\end{corollary}
Let us also point out some algebraic geometry content of the question studied in Lemmas~1 and 2 of \cite{ATL2010}.
\begin{remark}\label{remarkrationalnormal}
Let $V$ be an $\left(n-1\right)$-dimensional CES in $\mathbbm{C}^2\otimes\mathbbm{C}^n$. As proved in Lemma 1 of \cite{ATL2010}, the family of product vectors in $V^{\bot}$ has the form
\begin{equation}\label{familyprodvecrational}
\left(1,\alpha\right)\otimes A\left(1,\alpha,\ldots,\alpha^{n-1}\right)
\end{equation}
with $\alpha\in\mathbbm{C}$ and $A$ a non-singular $n\times n$ matrix. Such a family is a rational normal curve \cite[Example 18.8]{Harris}, as a general intersection of the Segre variety $\Sigma_{1,n-1}$ with an $n$-plane in $\mathbbm{P}^{2n-1}$ is.
\begin{proof}
Obviously, vectors of the form \eqref{familyprodvecrational} constitute a rational normal curve, cf. \cite[Example 18.8]{Harris}. A general intersection of $\Sigma_{1,n-1}$ with an $n$-plane in $\mathbbm{P}^{2n-1}$ will be nondegenerate and transverse. By Bezout's Theorem \cite[Theorem 18.3]{Harris}, it will be a nondegenate curve of degree $n$ in an $n$-plane. It is so because the degree of $\Sigma_{1,n-1}$ equals $n$. Consequently, the curve must be a rational normal curve, cf. Proposition 18.9 of \cite{Harris}. Thus we have explained our remark. However, it seems instructive to show why the intersection is transverse and nondegenerate, without repeating the whole argument of Lemma 1 in \cite{ATL2010}. Let us first discuss the issue of transversity. Take a $\phi\otimes\psi$ in $V^{\bot}$. As pointed out in \cite{ATL2010}, $V$ is spanned by $n-1$ product vectors $\Psi_i$ of Schmidt rank $2$, with some additional conditions they must fulfill. They can always be written as $\phi\otimes\psi^i_0+\phi^{\bot}\otimes\psi^i_1$, where $\innerpr{\phi}{\phi^{\bot}}=0$ and $\innerpr{\phi^{\bot}}{\phi^{\bot}}=1$, which uniquely defines the vector $\phi^{\bot}\subset\mathbbm{C}^2$, as well as forces $\psi^i_0$ and $\psi^i_1$ to be uniquely defined. Moreover, we know that $\left\{\psi^i_0\right\}_{i=1}^{n-1}$ and $\left\{\psi^i_1\right\}_{i=1}^{n-1}$ must be linearly independent. Otherwise, it is would be possible to find a product vector in $V$. As a conequence, $\psi$ is uniquely determined by the set of equations $\innerpr{\Psi_i}{\phi\otimes\psi}=0$ as the only vector in $\mathbbm{C}^n$ that fulfills $\innerpr{\psi^0_i}{\psi}$ for all $i=1,2,\ldots,n-1$. Motivated by the description of the tangent space $\mathbbm{T}_{\phi\otimes\psi}\left(\Sigma_{1,n-1}\right)$ to the Segre variety at $\phi\otimes\psi$, provided in Lemma \ref{lemmatangentSegre} from Section \ref{secprodvecPPT}, let us consider vectors of the form $\phi\otimes\psi'+\phi'\otimes\psi$ that belong to $V^{\bot}$. We can always assume $\phi'=\phi^{\bot}$. After we do so, $\psi'$ is determined by the set of equations $\innerpr{\Psi_i}{\phi\otimes\psi'+\phi^{\bot}\otimes\psi}=0\,\forall_{i=1,\ldots,n-1}$, which gives us $\innerpr{\psi^i_0}{\psi'}=-\innerpr{\psi^i_1}{\psi}$, $i=1,2,\ldots,n-1$. As we mentioned above, the vectors $\left\{\psi^i_0\right\}_{i=1}^{n-1}$, as well as $\left\{\psi^i_1\right\}_{i=1}^{n-1}$, are linearly independent, and $\psi$ is the only vector perpendicular to all the $\psi^i_0$'s. Therefore the equations $\innerpr{\psi^i_0}{\psi'}=-\innerpr{\psi^i_1}{\psi}$  have, up to a trivial scaling factor, a one-parameter family of solutions $\lambda\psi+\xi$, where $\xi$ is not proportional to $\psi$ and itself a solution to $\innerpr{\psi^i_0}{\xi}=-\innerpr{\psi^i_1}{\xi}$. Therefore the intersection of $\mathbbm{T}_{\phi\otimes\psi}\left(\Sigma_{1,n-1}\right)$ and $V^{\bot}$ has dimension $2$, or the projective dimension $1$. By a dimension counting argument practically identical to that presented in Lemma \ref{lemmageneralintersection}, we see that $V^{\bot}$ and $\Sigma_{1,n-1}$ intersect transversely. In order to prove that the intersection is a nondegenerate curve, it is sufficient to show that it spans $V^{\bot}$. Obviously, this is one of the statements of the Lemma \ref{lemmaAugusiak}, but let us show it in a somewhat more abstract way. First of all, it is easy to show, by a little modification of the argument above, that the equations $\innerpr{\Psi_i}{\phi\otimes\psi}$, $i=1,2,\ldots,n-1$, give exactly one solution for $\psi$ when $\phi$ is fixed. Therefore we have a family of product vectors in $V^{\bot}$, $\phi\otimes\psi\left(\phi\right)$. The precise form of the function $\psi\left(\phi\right)$, as in \eqref{remarkrationalnormal}, is not important to us at the moment. Just as the authors of \cite{ATL2010} did in the proof of their Lemma 1, it is fairly simple to show that the set $\left\{\psi\left(\phi\right)\right\}_{\phi\in\mathbbm{C}^2}$ must span $\mathbbm{C}^n$. Therefore it is possible to choose $n$ vectors $\phi_i\in\mathbbm{C}^2$ so that $\left\{\psi\left(\phi_i\right)\right\}_{i=1}^n$ is a basis of $\mathbbm{C}^n$. Choose another $\phi$, $\phi_{n+1}$, not equal to any of $\phi_i$ with $i=1,\ldots,n$. From the argument about transversity it is not difficult to see that $\psi\left(\phi_i\right)\neq\psi\left(\phi_j\right)$ for $\phi_i\neq\phi_j$. A simple argument then shows that the family of product vectors $\left\{\phi_i\otimes\psi\left(\phi_i\right)\right\}_{i=1}^{n+1}$ is linearly independent, and thus it must span $V^{\bot}$. This means nondegeneracy. In summary, the intersection of $V^{\bot}$ and $\Sigma_{1,n-1}$ is transverse and gives a nondegenerate curve of degree $n$ in an $n$-plane, which must be a rational normal curve. This is in full compliance with the specific form \eqref{familyprodvecrational}, obtained in \cite{ATL2010}.
\end{proof}
\end{remark}
We should also warn the reader of a too simple conjecture, which may be appealing because of the discussion above.
\begin{remark}\label{remarkwarning}
It is not true that any CES admits a gUPB of its orthogonal complement.
\begin{proof}
By \cite{Walgate08} we know that Completely Entangled Subspaces are generic among all subspaces of fixed dimension $k$, as long as $k$ is such that a $k$-dimensional CES can exist at all. Let us look at the case of $\mathbbm{C}^4\otimes\mathbbm{C}^4$. The corresponding maximum dimension of a CES is $16-4-4+1=9$, while the complementary dimension is $16-9=7$. Both of these numbers are clearly smaller or equal $9$. Choose a subspace $V\subset\mathbbm{C}^4\otimes\mathbbm{C}^4$ of dimension $9$ randomly. This is equivalent to choosing $V^{\bot}$ of dimension $7$ randomly. Since both $V$ and $V^{\bot}$ are chosen randomly, in the generic case neither of them contains a product vector, in accordance with the result of \cite{Walgate08}. One can confirm this numerically. 
\end{proof} 
\end{remark}

\section{Non-separable positive-partial-transpose states of lowest rank}\label{sec2}
The aim of the present section is to prove our main result, concerning positive-partial-transpose non-separable states of rank $4$ in $3\times 3$ systems. As indicated in \cite{LS2010}, they all seem to be possible to locally transform to projections onto the orthogonal complement to a subspace spanned by an orthogonal UPB. Thus, there is strong numerical evidence that they are all locally equivalent to bound entangled states of the form discussed in \cite{Bennett99}. In the following, we analytically prove that this is actually the case. Note that according to the results of \cite{HLVC2000}, four is the minimal rank for an entangled PPT state. Therefore it is correct to say that our theorem concerns non-separable PPT states of lowest rank.  Before we start with the proof, we need to discuss the concept of local equivalence, which will be fundamental to all that follows.

\subsection{The concept of local equivalence}\label{secequivalence}
Numerous questions of physical or mathematical origin need the proper identification of a group relevant to the problem in order to simplify the solution, or even to find it at all. The same is the case for the result we are going to obtain below. For PPT states, a natural group of symmetries should be of a product form, $\rho\mapsto\left(A\otimes B\right)^{\ast}\rho\left(A\otimes B\right)$, because all such transformations preserve the property of being PPT. In physical terms, they preserve the splitting of a composite system into subsystems, which is a highly desirable property. The remaining question is, what group should $A$ and $B$ belong to. When the amount of entanglement between the two subsystems is in question, a natural choice is $A$ and $B$ in the Unitary or Special Unitary group. Such transformations cannot change any measure of entanglement. However, if the aim is to classify PPT states with respect to the property of being extreme, being an edge state \cite{LKHC2001}, or the number and dimensionalities spanned by the product vectors in their kernels or ranges, $A$ and $B$ should most naturally belong to the General Linear or Special Linear group. There is no essential difference between the two latter choices. Since we are not interested in positive scaling factors in front of the states, we choose to work with the Special Linear group. This was also the approach so successfully used by the authors of \cite{LMO2006,LS2010,HHMS2011}. We should remark that, while a PPT state is transformed according to $\rho\mapsto\left(A\otimes B\right)^{\ast}\rho\left(A\otimes B\right)$, the product vectors in its kernel and its range undergo the following transformation, $\phi\otimes\psi\mapsto\left(A^{-1}\otimes B^{-1}\right)\phi\otimes\psi$. Conversely, a transformation $\phi\otimes\psi\mapsto\left(A\otimes B\right)\phi\otimes\psi$ forces a change of $\rho$ into $\left(A^{-1}\otimes B^{-1}\right)^{\ast}\rho\left(A^{-1}\otimes B^{-1}\right)$. It is these kind of transformations we will have in mind when we talk about ``local equivalence'', ``local $\SL$ equivalence'' or ``$\SLtt$ equivalence'' in the following sections. Any similar terms, even not listed here, will also refer to precisely the same situation. Nevertheless, when product vectors in the kernel of a PPT state $\rho$ are in question, it is more convenient to look at them as rays, points in the projective space. In such case, it is also more accurate to refer to the projectivisation of the group $\SLtt$, namely to $\PSLtt$.  In simple words, we may multiply vectors $\left\{\phi_1\otimes\psi_1,\phi_2\otimes\psi_2,\ldots\right\}\subset\kernel{\rho}$ by arbitrary individual factors, and they will remain elements of the kernel of $\rho$. Together with the previously introduced product $\SLtt$ symmetries, we have a group of transformations that is most properly described as $\PSLtt$. Note that the use of this term is motivated mainly by the possibility to avoid excessive comments about constant factors in front of the product vectors in $\kernel{\rho}$. We are fully legitimate to use the previously introduced name ``local equivalence'' also for the $\PSLtt$ transformations we just described because constant factors are completely irrelevant to $\rho$ itself.

The ultimate reason for using equivalences of the form described above will be the simplicity of our main result, a characterization theorem that we are going to obtain in Section \ref{secmainresult}. The equivalence classes under $\SLtt$ of non-separable PPT states of rank $4$ in $3\times 3$ systems turn out to be parametrized by just four real, positive numbers. Moreover, each class has a representative which is a projection onto a Completely Entangled Subspace complementary to a $3\times 3$ orthogonal UBP. This is a surprising result for which strong numerical evidence was provided in \cite{LS2010} and later supported by certain analytical results of \cite{HHMS2011}.

\subsection{Outline of the proof}\label{secoutline}
The proof is not excessively complicated, but it needs a considerable amount of work. It also consists of a number of steps which do not seem easy to merge. In order to simplify the reading, we start with a list of building blocks. We will elaborate on each of them in the following sections.
\begin{enumerate}
\item The kernel of a rank four PPT state $\rho$ must intersect the Segre variety in a transverse way. In particular, according to the Bezout's Theorem, the intersection must consist of exactly six points.
\item The product vectors in the kernel of a rank $4$ PPT state in the $3\times 3$ case span the kernel. As a result, they must be a generalized UPB. There cannot exist a product vector orthogonal to all of them.
\item A generalized UPB in the $3\times 3$ case is locally equivalent to an orthogonal one if and only if certain invariants $s_1,\ldots,s_4$, introduced in \cite{LS2010}, are all positive, possibly after the vectors are permuted.
\item A generalized UPB in a $3\times 3$ system is contained in a kernel of some rank four PPT state if and only if the corresponding values of $s_1,\ldots,s_4$ are positive, possibly after the vectors are permuted. Moreover, in such case the PPT state in question is uniquely determined.
\end{enumerate}
The final conclusion from the facts mentioned in items $\left(1\right)-\left(4\right)$ is that the only non-separable PPT states of rank $4$ in $3\times 3$ systems are local transforms of projections onto orthogonal complements of orthogonal pentagram-type Unextendible Product Bases.

\subsection{Product vectors in the kernel of a PPT state}\label{secprodvecPPT}
The present section elaborates on item $\left(1\right)$ in the list given above and on related topics. Let us start with an elementary fact
\begin{lemma}\label{lemmaconj}
A product vector $\wektor{\phi\otimes\psi}$ is in the kernel of a PPT state $\rho$ if and only if the partially conjugated states $\wektor{\phi^{\ast}\otimes\psi}$ and $\wektor{\phi\otimes\psi^{\ast}}$ are in the kernels of $\rho^{T_1}$ and $\rho^{T_2}$, respectively.
\begin{proof}
It simply follows from the identity $\innerpr{\phi\otimes\psi}{\rho\left(\phi\otimes\psi\right)}=\innerpr{\phi\otimes\psi^{\ast}}{\rho^{T_2}\left(\phi\otimes\psi^{\ast}\right)}$ and $\innerpr{\phi\otimes\psi}{\rho\left(\phi\otimes\psi\right)}=\innerpr{\phi^{\ast}\otimes\psi}{\rho^{T_1}\left(\phi^{\ast}\otimes\psi\right)}$, by the positivity of $\rho$, $\rho^{T_1}$ and $\rho^{T_2}$.
\end{proof}
\end{lemma}
In the above lemma, we did not assume anything about the dimensionality of the system. Neither we do it in the following.
\begin{lemma}\label{lemmaprodPPT}
Assume that a product vector $\wektor{\phi\otimes\psi}$ is in the kernel of a PPT state $\rho$. In such case
\begin{equation}\label{condPPTproduct}
\innerpr{\phi'\otimes\psi}{\rho\left(\phi\otimes\psi'\right)}=\innerpr{\phi\otimes\psi'}{\rho\left(\phi'\otimes\psi\right)}=0\quad\forall_{\phi',\psi'}
\end{equation}
\begin{proof} Since $\rho\left(\wektor{\phi\otimes\psi}\right)=0$, we know from Lemma \ref{lemmaconj} that $\rho^{T_1}\left(\wektor{\phi^{\ast}\otimes\psi}\right)=0$, which obviously implies $\innerpr{\phi'^{\ast}\otimes\psi'}{\rho^{T_1}\left(\phi^{\ast}\otimes\psi\right)}=\innerpr{\phi\otimes\psi'}{\rho\left(\phi'\otimes\psi\right)}=0$. This is the first equality in \eqref{condPPTproduct}. The second one can be obtained in a similar way.
\end{proof}
\end{lemma}
Our next lemma applies specifically to the $3\times n$ case. For reference concerning edge PPT states, check \cite{LKHC2001}. 
\begin{lemma}\label{lemma3x3twoproducts}
Assume that both $\phi\otimes\psi$ and $\wektor{\phi'\otimes\psi}$, with $\phi$, $\phi'$ in $\mathbbm{C}^3$ and $\psi$ in $\mathbbm{C}^n$, $\phi\neq\phi'$, belong to the kernel of a PPT state $\rho$, acting on $\mathbbm{C}^3\otimes\mathbbm{C}^3$. The state $\rho$ is either supported on a $3\times\left(n-1\right)$ or smaller subspace, or it can be written as $\rho'+\lambda\proj{\phi''\otimes\xi}{\phi''\otimes\xi}$ for some $\lambda>0$, $\xi\in\mathbbm{C}^n$, $\phi''\in\mathbbm{C}^3$ linearly independent of $\phi$ and $\phi'$, and a PPT state $\rho'$, supported on a $3\times\left(n-1\right)$ or smaller subspace. Moreover $\rank{\rho'}=\rank{\rho}-1$ and $\rank{\left(\rho'\right)^{T_1}}=\rank{\rho^{T_1}}-1$.
In a situation when the reduction is possible, the state $\rho$ is not an edge PPT state. In particular, $\rho$ is not an extreme and non-separable PPT state. 
\begin{proof}
Let us assume that the product states $\phi'\otimes\psi$ and $\wektor{\phi_2\otimes\psi}$ belong to the kernel of $\rho$. Let $A$ be an $\SLt$ transformation that brings $e_1,e_2\subset\mathbbm{C}^3$ to $\phi_1$ and $\phi_2$. A little inspection shows that Lemmas~1 and 2 of \cite{HLVC2000} can be applied to $\tilde\rho:=\left(A\otimes\mathbbm{1}\right)^{\ast}\rho\left(A\otimes\mathbbm{1}\right)$. Consequently, we see that either $\rho$ is supported on a $3\times\left(n-1\right)$ or smaller space, or the assertion of Lemma~2 of \cite{HLVC2000} tells us that $\tilde\rho=\rho_1+\lambda\proj{e_3\otimes\xi}{e_3\otimes\xi}$ for some $\xi\in\mathbbm{C}^n$, and moreover, $\rho_1$ is a PPT state supported on a $3\times\left(n-1\right)$ or smaller subspace, with $\rank{\rho_1}=\rank{\rho}-1$ and $\rank{\rho_1^{T_1}}=\rank{\rho^{T_1}}-1$. We have $\rho=\left(A^{-1}\otimes\mathbbm{1}\right)^{\ast}\tilde\rho\left(A^{-1}\otimes\mathbbm{1}\right)=\rho'+\lambda\proj{\phi'''\otimes\xi}{\phi'''\otimes\xi}$, where  $\phi'''=A^{-1}e_3$ and $\rho'=\left(A^{-1}\otimes\mathbbm{1}\right)^{\ast}\rho_1\left(A^{-1}\otimes\mathbbm{1}\right)$. The states $\rho'$ and $\left(\rho'\right)^{T_1}$ still have their ranks reduced by one with respect to the ranks of $\rho$ and $\rho^{T_1}$, respectively. The subspaces on which they are supported are of the same type as for $\rho_1$, hence $3\times\left(n-1\right)$ or smaller. The statement that $\rho$ is not an edge state simply follows because $\proj{\phi'''\otimes\xi}{\phi'''\otimes\xi}$ is in $\range{\rho}$ while its partial conjugation is in $\range{\rho^{T_1}}$.
\end{proof}
\end{lemma}
The following result reduces a more general case to the situation considered above. However, this time we assume $n=3$.
\begin{lemma}\label{lemmaSR2}
Let $\wektor{\phi\otimes\psi}\in\mathbbm{C}^3\otimes\mathbbm{C}^3$ be an element of a PPT state $\rho$, acting on $\mathbbm{C}^3\otimes\mathbbm{C}^3$. There cannot exist a nonzero vector $\wektor{\phi\otimes\psi'}+\wektor{\phi'\otimes\psi}$, with $\phi'\neq\phi$ or $\psi'\neq\psi$, in the kernel of $\rho$, unless one of the following is true: i) $\rho=\rho'+\lambda\proj{\zeta\otimes\xi}{\zeta\otimes\xi}$ for $\lambda>0$, $\xi,\zeta\in\mathbbm{C}^3$ and $\rho'$ a PPT state supported on a $2\times 3$ or smaller subspace with $\rank{\rho'}=\rank{\rho}-1$ and  $\rank{\left(\rho'\right)^{T_1}}=\rank{\left(\rho\right)^{T_1}}-1$ or ii) $\rho$ is supported on a $2\times 3$ or smaller subspace itself.
\begin{proof}
Assume that there is a state of the form $\wektor{\phi\otimes\psi'}+\wektor{\phi'\otimes\psi}$ in the kernel of $\rho$. This is equivalent to saying that $\innerpr{\phi\otimes\psi'+\phi'\otimes\psi}{\rho\left(\phi\otimes\psi'+\phi'\otimes\psi\right)}=0$. The inner product factorizes as
\begin{equation}\label{factorinnerpr}
\innerpr{\phi\otimes\psi'}{\rho\left(\phi\otimes\psi'\right)}+\innerpr{\phi\otimes\psi'}{\rho\left(\phi'\otimes\psi\right)}+\innerpr{\phi'\otimes\psi}{\rho\left(\phi\otimes\psi'\right)}+\innerpr{\phi'\otimes\psi}{\rho\left(\phi'\otimes\psi\right)}\nonumber
\end{equation}
The two factors in the middle vanish according to Lemma \ref{lemmaprodPPT}, while the two remaining factors are nonnegative as a consequence of positivity of $\rho$. Therefore, the only possibility for the above expression to vanish is when $\innerpr{\phi\otimes\psi'}{\rho\left(\phi\otimes\psi'\right)}=0$ and $\innerpr{\phi'\otimes\psi}{\rho\left(\phi'\otimes\psi\right)}=0$. This in turn means that $\rho\left(\wektor{\phi'\otimes\psi}\right)=0$ and $\rho\left(\wektor{\phi\otimes\psi'}\right)=0$. According to our assumptions, at least one of these equalities is nontrivial (i.e. $\phi'\neq 0$ or $\psi'\neq 0$). Lemma \ref{lemma3x3twoproducts} can be applied.
\end{proof}
\end{lemma}
The importance of Lemma \ref{lemmaSR2} is evident if we realize that the tangent space to the Segre variety, or to the set of product states at a point $\wektor{\phi\otimes\psi}$, consists precisely of the vectors  of the form considered above. We have
\begin{lemma}\label{lemmatangentSegre}
Elements of the tangent space to the Segre variety, or to the set of product vectors at a point $\wektor{\phi\otimes\psi}$, are of the form
\begin{equation}\label{formtangent}
\wektor{\phi\otimes\psi'}+\wektor{\phi'\otimes\psi},
\end{equation} 
with $\psi'$ and $\phi'$ arbitrary.
\begin{proof} A heuristic proof may consist in writing $\wektor{\left(\phi+\delta\phi\right)\otimes\left(\psi+\delta\psi\right)}\approx\wektor{\phi\otimes\psi}+\wektor{\delta\phi\otimes\psi}+\wektor{\phi\otimes\delta\psi}$, where the approximate equality holds to the first order. A slightly more rigorous one can be found in Example 14.16 of \cite{Harris}.
\end{proof}
\end{lemma}
Next, we specify the rank of $\rho$ to be $4$ and keep the assumption that $\rho$ acts on $\mathbbm{C}^3\otimes\mathbbm{C}^3$. Thus the kernel of $\rho$ is of dimension $5$, which is the smallest number $d$ such that a $d$-dimensional linear subspace must intersect the set of product vectors in $\mathbbm{C}^3\otimes\mathbbm{C}^3$, cf. e.g. \cite{Partha04}. Following Lemmas \ref{lemmaSR2} and \ref{lemmatangentSegre}, we can show that the nonempty intersection is generic in the sense of Bezout's theorem \cite[Theorem 18.3]{Harris} and thus it consists of exactly six points.
\begin{lemma}\label{lemmageneralintersection}
Let $\rho$ be a non-separable PPT state of rank $4$ acting on $\mathbbm{C}^3\otimes\mathbbm{C}^3$. The intersection between the respective Segre variety and the five-dimensional kernel of $\rho$ is transverse at every point. There are exactly six product vectors in the kernel of $\rho$.
\begin{proof}
The preceeding lemmas and the arguments included in their proofs imply directly.
Let us take $\wektor{\phi\otimes\psi}\in\kernel{\rho}$. As we mentioned above, such a vector exists \cite{Partha04,Cubitt07} by a dimensionality argument  for projective varieties. We easily see from Lemma \ref{lemmatangentSegre} that the dimension of the tangent space $\mathbbm{T}_{\wektor{\phi\otimes\psi}}\left(\Sigma_{2,2}\right)$ to the Segre variety at $\wektor{\phi\otimes\psi}$ is $5$, and thus the projective dimension is $4$. Being more explicit, any vector of the form $\wektor{\phi\otimes\psi'}+\wektor{\phi'\otimes\psi}$ can be written as $\lambda\phi\otimes\psi+\sum_{i=1}^2\xi_i\phi\otimes\psi_i+\sum_{j=1}^2\zeta_j\phi_j\otimes\psi$, where $\left\{\phi,\phi_1,\phi_2\right\}$ and $\left\{\psi,\psi_1,\psi_2\right\}$ are two sets of  three linearly independent vectors in $\mathbbm{C}^3$ and $x_i$, $\zeta_j$ are arbitrary complex coefficients. From Lemmas \ref{lemmaSR2} and \ref{lemmatangentSegre} we know that the only vector in the intersection of $\kernel{\rho}$ and $\mathbbm{T}_{\wektor{\phi\otimes\psi}}\left(\Sigma_{2,2}\right)$ is $\wektor{\phi\otimes\psi}$ itself. It must be so, because otherwise we could reduce the rank of $\rho$ by subtracting a projection onto a product state. After the reduction, we would be left with a PPT state of rank $3$. However, all such PPT states are separable according to \cite{HLVC2000}, and $\rho$ would have to be separable as well. The other option is that $\rho$ could be supported on $\mathbbm{C}^2\otimes\mathbbm{C}^3$ or even a less dimensional space itself. But then it is well-known that $\rho$ is separable as as consequence of being PPT \cite{2x3separable}. In either case, we get a contradiction with the assumption that $\rho$ is non-separable. Therefore, $\wektor{\phi\otimes\psi}$ must be, up to a scalar factor, the only element of the intersection between $\kernel{\rho}$ and $\mathbbm{T}_{\wektor{\phi\otimes\psi}}\left(\Sigma_{2,2}\right)$. Consequently, the dimension of $\kernel{\rho}+\mathbbm{T}_{\wektor{\phi\otimes\psi}}\left(\Sigma_{2,2}\right)$ equals $5+5-1=9$, and the projective dimension is $9-1=8$. This equals the projective dimension of $\mathbbm{T}_{\wektor{\phi\otimes\psi}}\left(\mathbbm{P}^8\right)$, or simpler, the dimension of the projective space $\mathbbm{P}^8$. In other words, $\kernel{\rho}$ and $\mathbbm{T}_{\wektor{\phi\otimes\psi}}\left(\Sigma_{2,2}\right)$ span $\mathbbm{T}_{\wektor{\phi\otimes\psi}}\left(\mathbbm{P}^8\right)$, which is equivalent to saying that the intersection between $\kernel{\rho}$ and the Segre variety is transverse at $\wektor{\phi\otimes\psi}$. Since we did not make any additional assumptions about $\wektor{\phi\otimes\psi}$ apart from that it belongs to the intersection, we see that the intersection is transverse at every point. Therefore Bezout's theorem applies. The fact that there are exactly six points in the intersection follows because the degree of the Segre variety $\Sigma_{2,2}$ is six \cite[Example 18.15]{Harris}. 
\end{proof}
\end{lemma}
In summary, in the present section we have shown that a non-separable rank $4$ PPT state in a $3\times 3$ system must have exactly six vectors in its kernel. This is in full argeement with an assertion of \cite{LS2010n}. It should be noticed that, as a part of the proof of the above lemma, we have shown that non-separable PPT states of rank $4$ in $3\times 3$ systems  are edge states.  Thus, Lemmas \ref{lemma3x3twoproducts} and \ref{lemmaSR2} can be directly applied. We will frequently use them in the following section. 

\subsection{Product vectors in the kernel must be a gUPB}\label{secgUPB}
We already know that the number of product vectors in the kernel of a rank $4$ non-separable PPT state of a $3\times 3$ system is six. In the following, we discuss more specific properties of the set of six product vectors. Let us denote them with $\phi_i\otimes\psi_i$, $i=1,2,\ldots,6$. It turns out that, up to local equivalence, five of them can always be brought to a special form, which has only four real parameters, the numbers $s_1,\ldots,s_4$ introduced in \cite{LS2010}. It then follows that the vectors $\phi_i\otimes\psi_i$, if they belong to the kernel of a rank $4$ PPT state, must span a five-dimensional subspace. Thus they span the kernel.

In order to prove our assertion, first observe that $\phi_i\neq\phi_j$ for $i\neq j$, and thus they must span at least a two-dimensional subspace of $\mathbbm{C}^3$. Similarly for the $\psi$'s. Let us assume first that one of the set $\left\{\phi_i\right\}_{i=1}^6$ or $\left\{\psi_j\right\}_{j=1}^6$ spans a two-dimensional subspace. We may, for example, try to assume this about $\left\{\phi_i\right\}_{i=1}^6$. Up to $\PSLt$ transformations, we have
\begin{equation}\label{twodimphi}
\left[\begin{array}{cccccc}\phi_1 & \phi_2 & \phi_3 & \phi_4 & \phi_5 & \phi_6\end{array}\right]=\left[\begin{array}{cccccc}
1&0&1&1&1&1\\
0&1&1&p&q&r\\
0&0&0&0&0&0
\end{array}
\right],
\end{equation}
where $p$, $q$, $r$ are all different and different from $0$ and $1$. When writing \eqref{twodimphi}, we used the fact that there is no pair of identical vectors in $\left\{\phi_i\right\}_{i=1}^6$.  Up to local transformations, we have $\psi_1=e_1$ and $\psi_2=e_2$. As for the other vectors $\psi$, we use the following notation, $\psi_i=\left[\begin{array}{ccc}\psi_{1i}&\psi_{2i}&\psi_{3i}\end{array}\right]$, $i=3,4,5,6$. We also introduce coordinates $\omega^{ij}$ for general vectors $\wektor{\omega}=\sum_{i,j}\omega^{ij}e_i\otimes e_j$ in $\mathbbm{C}^3\otimes\mathbbm{C}^3$. Our aim is to show that there exists a linear combination of the vectors $\phi_i\otimes\psi_i$ of the form  $\wektor{\phi\otimes\psi'}+\wektor{\phi'\otimes\psi}$ from Lemma \ref{lemmatangentSegre}. This will lead us to a contradiction and show that $\phi_i$'s cannot be as in \eqref{twodimphi}, and must span $\mathbbm{C}^3$. An analogous conclusion for $\psi$'s will be immediate.

Let us first observe that $\psi_{3i}\neq 0$ for all $i\in\left\{3,4,5,6\right\}$. Otherwise, we would have three product vectors supported on $\linspan{e_1,e_2}\otimes\linspan{e_1,e_2}$. Up to local equivalence, they would be of the form $e_1\otimes e_1$, $e_2\otimes e_2$ and $\left(e_1+e_2\right)\otimes\left(e_1+e_2\right)$. In such case, $e_1\otimes e_2+e_2\otimes e_1=\left(e_1+e_2\right)\otimes\left(e_1+e_2\right)-e_1\otimes e_1-e_2\otimes e_2$ would be in the kernel of $\rho$, which contradicts Lemma \ref{lemmaSR2}. Therefore we must have $\psi_{3i}\neq 0$ for all $i$. Let us choose $\alpha$ and $\beta$ so that $\alpha\psi_{33}+\beta p\psi_{34}=0$. The vector $\alpha\phi_3\otimes\psi_3+\beta\phi_4\otimes\phi_4$ has a vanishing coordinate $\omega^{23}=\alpha\psi_{33}+\beta p\psi_{34}$ and a non-vanishing coordinate $\omega^{13}=\alpha\psi_{33}+\beta\psi_{34}$ (remember that $p\neq 1$). By subtracting $e_2\otimes e_2$ times $\alpha\psi_{23}+\beta q\psi_{34}$, we can cancel the $\omega^{22}$ coordinate, and similarly cancel $\omega^{11}$ by subtracting $\alpha\psi_{13}+\beta \psi_{14}$ times $e_1\otimes e_1$. In the end, we see that a vector of the form $\omega^{21} e_2\otimes e_1+\omega^{12} e_1\otimes e_2+\omega^{13}e_1\otimes e_3$ with $\omega^{13}\neq 0$ is in the kernel of $\rho$. But this contradicts Lemma \ref{lemmaSR2}. In summary, the vectors $\phi_i$ cannot be brought to the form \eqref{twodimphi}, or in other words, they span $\mathbbm{C}^3$. Obviously, the same is true for the set $\left\{\psi_i\right\}_{i=1}^6$. A more careful analysis of the above argument leads to even stronger conclusions. Firstly, an assumption that there exist three vectors $\phi_i\otimes\psi_i$ supported on a $2\times 2$ dimensional subspace lead us to a contradiction. Therefore we have the following
\begin{lemma}\label{lemmanotwotwo}
Let $\left\{\phi_i\otimes\psi_i\right\}_{i=1}^6$ be defined as  above. For any triple $\left\{\phi_{i_j}\otimes\psi_{i_j}\right\}_{j=1}^3\subset\left\{\phi_i\otimes\psi_i\right\}_{i=1}^6$, at least one of the sets of vectors $\left\{\phi_{i_j}\right\}_{j=1}^3$ or $\left\{\psi_{i_j}\right\}_{j=1}^3$ spans $\mathbbm{C}^3$.
\end{lemma}
Moreover, we only needed four product vectors with $\phi$'s as in \eqref{twodimphi} to arrive at a contradiction with Lemma \ref{lemmaSR2}. As a consequence, we have
\begin{lemma}\label{lemmaquadruples}
For any quadruple $\left\{\phi_{i_j}\otimes\psi_{i_j}\right\}_{j=1}^4\subset\left\{\phi_i\otimes\psi_i\right\}_{i=1}^6$, both the sets of vectors $\left\{\phi_{i_j}\right\}_{j=1}^4$ and $\left\{\psi_{i_j}\right\}_{j=1}^4$ span $\mathbbm{C}^3$.
\end{lemma}
As an immediate consequence of Lemma \ref{lemmanotwotwo}, there exists a set of three linearly independent vectors in $\left\{\phi_i\right\}_{i=1}^6$. With no loss of generality, we may assume that $\left\{\phi_1,\phi_2,\phi_6\right\}$ is a linearly independent set. After a $\PSLt$ transformation, $\phi_1=e_1$, $\phi_2=e_2$ and $\phi_6=e_3$. There are in principle two possibilities concerning the remaining vectors $\phi_3$, $\phi_4$ and $\phi_5$. Either one of them is of the form $\left[\begin{array}{ccc}x&y&z\end{array}\right]$ with $xyz\neq 0$, or all of them have exactly one coordinate equal to zero. Two vanishing coordinates in a single vector cannot occur because there is no pair of identical vectors among $\phi_1,\ldots,\phi_6$. Moreover, according to Lemma \ref{lemmaquadruples}, the zeros must occur in different places in $\phi_3$, $\phi_4$ and $\phi_5$. Up to $\PSLt$ transformations and permuting the vectors, we may assume that $\phi_3=\left[\begin{array}{ccc}x&0&1\end{array}\right]$,  $\phi_4=\left[\begin{array}{ccc}0&1&z\end{array}\right]$, $\phi_5=\left[\begin{array}{ccc}1&y&0\end{array}\right]$ with $x$, $y$, $z$ all different from $0$. But then, write the coordinate matrix for $\left\{\phi_1,\phi_2,\phi_3,\phi_4\right\}$,
\begin{equation}\label{coordmatrix2345}
\left[\begin{array}{cccc}\phi_1 & \phi_2 & \phi_3 & \phi_4\end{array}\right]=\left[\begin{array}{cccc}
1&0&x&0\\
0&1&0&1\\
0&0&1&z
\end{array}
\right]
\end{equation} 
It is easy to check that all the $3\times 3$ minors in \eqref{coordmatrix2345} are non-vanishing. In other words, any triple of vectors in $\left\{\phi_1,\phi_2,\phi_3,\phi_4\right\}$ spans $\mathbbm{C}^3$. The corresponding vectors $\psi_1$, $\psi_2$, $\psi_3$ and $\psi_4$ may or may not have all triples linearly independent. If all the triples span $\mathbbm{C}^3$, we can simultaneously, by using a $\PSLtt$ transformation, bring $\left\{\phi_1,\phi_2,\phi_3,\phi_4\right\}$ and $\left\{\psi_1,\psi_2,\psi_3,\psi_4\right\}$ to the form
\begin{equation}\label{formtwoindeptriples}
\left[\begin{array}{cccc}\phi_1 & \phi_2 & \phi_3 & \phi_4\end{array}\right]=\left[\begin{array}{cccc}\psi_1 & \psi_2 & \psi_3 & \psi_4\end{array}\right]=\left[\begin{array}{cccc}
1&0&0&1\\
0&1&0&1\\
0&0&1&1
\end{array}
\right]
\end{equation} 
By adding a fifth product vector, say $\phi_5\otimes\psi_5$, we get, up to local transformation and relabelling the vectors $\phi_i\otimes\psi_i$,
\begin{equation}\label{formtwoindeptriples2}
\left[\begin{array}{ccccc}\phi_1&\phi_2&\phi_3&\phi_4&\phi_5\\
\hline
\psi_1&\psi_2&\psi_3&\psi_4&\psi_5
\end{array}
\right]=\left[\begin{array}{ccccc}
1&0&0&1&1\\
0&1&0&1&p\\
0&0&1&1&q\\
\hline
1&0&0&1&1\\
0&1&0&1&r\\
0&0&1&1&s
\end{array}
\right],
\end{equation}
where $p,q,r,s$ are some complex numbers. We should remark that the possibility to have $1$ in the first coordinate of $\phi_5$ and $\psi_5$ follows because there must exist $i\in\left\{1,2,3\right\}$ such that $\phi_{i5}\psi_{i5}\neq 0$. Otherwise, $\phi_5$ or $\psi_5$ would have to be proportional to $e_i$ for some $i\in\left\{1,2,3\right\}$.
 
If not all triples in $\left\{\psi_1,\psi_2,\psi_3,\psi_4\right\}$ are linearly independent, it is still possible, according to Lemma~\ref{lemmaquadruples}, to find a linearly independent triple among them. Without loss of generality, we may assume that the triple is $\left\{\psi_1,\psi_2,\psi_3\right\}$. By an identical argument as for the $\phi$'s, we know that there is a vector $\psi_i$, $i\in\left\{5,6\right\}$ such that $\left\{\psi_1,\psi_2,\psi_3,\psi_i\right\}$ have all triples linearly independent. Without loss of generality, we may assume that $\phi_i=\phi_5$. This time, a local transformation and possible relabelling brings the product vectors $\phi_i\otimes\psi_i$ with $i=1,2,\ldots,5$ to the form
\begin{equation}\label{formtwoindeptriples3}
\left[\begin{array}{ccccc}\phi_1&\phi_2&\phi_3&\phi_4&\phi_5\\
\hline
\psi_1&\psi_2&\psi_3&\psi_4&\psi_5
\end{array}
\right]=\left[\begin{array}{ccccc}
1&0&0&1&1\\
0&1&0&1&p\\
0&0&1&1&q\\
\hline
1&0&0&1&1\\
0&1&0&r&1\\
0&0&1&s&1
\end{array}
\right].
\end{equation}

To make a final touch to this section, we need to show that product vectors of the form \eqref{formtwoindeptriples2} or \eqref{formtwoindeptriples3} are linearly independent if they are five in number, and thus they span the five-dimensional kernel of $\rho$. We will also show that they constitute a minimal gUPB, and that the parameters $p,q,r,s$ have to be real when the vectors are in the kernel of a PPT state.

Let us use the notation $\left[\begin{array}{ccccccccc}\omega^{11}&\omega^{12}&\omega^{13}&\omega^{21}&\omega^{22}&\omega^{23}&\omega^{31}&\omega^{32}&\omega^{33}\end{array}\right]$ for vectors $\omega=\sum_{i,j}\omega^{ij}e_i\otimes e_j$ in $\mathbbm{C}^3\otimes\mathbbm{C}^3$. In the case \eqref{formtwoindeptriples2}, we have
\begin{equation}\label{sit1coordinates}
\left[\begin{array}{c}
\phi_1\otimes\psi_1\\
\phi_2\otimes\psi_2\\
\phi_3\otimes\psi_3\\
\phi_4\otimes\psi_4\\
\phi_5\otimes\psi_5\\
\end{array}
\right]=
\left[\begin{array}{ccccccccc}
1&0&0&0&0&0&0&0&0\\
0&0&0&0&1&0&0&0&0\\
0&0&0&0&0&0&0&0&1\\
1&1&1&1&1&1&1&1&1\\
1&r&s&p&pr&ps&q&qr&qs\end{array}
\right]
\end{equation}
In the case \eqref{formtwoindeptriples3}, the coordinates of the product vectors are the following,
\begin{equation}\label{sit2coordinates}
\left[\begin{array}{c}
\phi_1\otimes\psi_1\\
\phi_2\otimes\psi_2\\
\phi_3\otimes\psi_3\\
\phi_4\otimes\psi_4\\
\phi_5\otimes\psi_5\\
\end{array}
\right]=
\left[\begin{array}{ccccccccc}
1&0&0&0&0&0&0&0&0\\
0&0&0&0&1&0&0&0&0\\
0&0&0&0&0&0&0&0&1\\
1&r&s&1&r&s&1&r&s\\
1&1&1&p&p&p&q&q&q\end{array}
\right]
\end{equation}
It is an elementary exercise to check that the matrices on the right-hand side of \eqref{sit1coordinates} and \eqref{sit2coordinates} are of rank $5$ for all choices of $p,q,r,s$, with the only exception of $p=q=r=s=1$. But the last possibility is excluded because it implies $\phi_4\otimes\psi_4=\phi_5\otimes\psi_5$.

Next, we can show that the vectors $\phi_i\otimes\psi_i$ with $i=1,2,\ldots,5$, chosen as above, constitute a general Unextendible Product Basis.  In order to prove it, let us first show that the rank of $\rho^{T_1}$ has to be $4$. 
\begin{proposition}\label{propequalranks}
Let $\rho$ be a non-separable PPT state of rank $4$ acting on $\mathbbm{C}^3\otimes\mathbbm{C}^3$. The rank of $\rho^{T_1}$ is also $4$.
\begin{proof}
If $\rho$ is non-separable, we know by the above argument that the product vectors $\left\{\phi_i\otimes\psi_i\right\}_{i=1}^6$ in the kernel of $\rho$ span a five-dimensional subspace, which is the kernel itself. Moreover, five of them are, up to local transformations, of the form \eqref{formtwoindeptriples2} or \eqref{formtwoindeptriples3}. But this implies that the corresponding product vectors in the kernel of $\rho^{T_1}$, which are $\phi_i^{\ast}\otimes\psi_i$ according to Lemma \ref{lemmaconj}, can also be brought to the form \eqref{formtwoindeptriples2} or \eqref{formtwoindeptriples3}. To be more explicit, if a local transformation $A\otimes B$ brings the vectors $\phi_i\otimes\psi_i$ with $i=1,2,\ldots,5$ to the form \eqref{formtwoindeptriples2} or \eqref{formtwoindeptriples3}, $A^{\ast}\otimes B$ does the same to the partial conjugations $\phi_i^{\ast}\otimes\psi_i$. The only difference is that $p$ and $q$ change into $p^{\ast}$ and $q^{\ast}$ in \eqref{formtwoindeptriples2} or \eqref{formtwoindeptriples3}. But this does not change the conclusion about the dimensionality of the subspace spanned by vectors of the form \eqref{formtwoindeptriples2} or \eqref{formtwoindeptriples3}. As a consequence, the product vectors in the kernel of $\rho^{T_1}$ span at least a five-dimensional subspace. Thus the kernel of $\rho^{T_1}$ is at least five-dimensional. If it had higher dimension, the rank of $\rho^{T_1}$ would be lower or equal $3$, which is, according to \cite{HLVC2000}, impossible for non-separable $\rho$. Therefore, the dimension of the kernel equals $5$, and the rank of $\rho^{T_1}$ is $4$.
\end{proof}
\end{proposition}
There exist  separable states $\rho$ of rank $4$ in $3\times 3$ systems that have the rank of $\rho^{T_1}$ different from $4$. However, our next proposition shows that if $\rho$ is supported on $\mathbbm{C}^3\otimes\mathbbm{C}^3$ and it cannot be written as $\rho'+\lambda\proj{\zeta\otimes\xi}{\zeta\otimes\xi}$ with $\lambda>0$ and $\rho'$ supported on a $2\times 2$ subspace (cf. Figure 4 in \cite{LS2010n}), the rank of $\rho^{T_1}$ is also $4$.
\begin{proposition}\label{propranksseparable}
Let $\rho$ be a separable state of rank $4$ supported on $\mathbbm{C}^3\otimes\mathbbm{C}^3$, which cannot be written as $\rho'+\lambda\proj{\zeta\otimes\xi}{\zeta\otimes\xi}$ with $\lambda>0$ and $\rho'$ supported on a $2\times 2$ subspace of $\mathbbm{C}^3\otimes\mathbbm{C}^3$. The rank of $\rho^{T_1}$ is also $4$.
\begin{proof} 
First, we should remark that $\rank{\rho^{T_1}}=\rank{\rho^{T_2}}$. This fact will be important for some parts of the proof, although never explicitly refered to. The main idea that we are going to use is that the argument preceeding formulas \eqref{formtwoindeptriples2} and \eqref{formtwoindeptriples3} works for separable states as well, provided that they cannot be reduced according to Lemma \ref{lemmaSR2}. In other words, the argument works when the kernel of a PPT state in question does intersect the Segre variety in a transverse way. Thus, if a reduction according to Lemma \ref{lemmaSR2} is not possible for a separable state $\rho$, we have vectors of the form \eqref{formtwoindeptriples2} or \eqref{formtwoindeptriples3} in $\kernel{\rho}$, and they span a five-dimensional space. This is also the dimensionality of the subspace spanned by their partial conjugates, which are in $\kernel{\rho^{T_1}}$. Therefore the rank of $\rho^{T_1}$ is not bigger than $4$. If it was less than four, the intersection between $\kernel{\rho^{T_1}}$ and the Segre variety $\Sigma_{2,2}$ would be more than zero-dimensional, according to the Projective Dimension Theorem \cite[Theorem 7.2]{Hartshorne}. But this contradicts the fact that there are only a finite number of product vectors in $\kernel{\rho^{T_1}}$ (equal to $\phi^{\ast}\otimes\psi$ for all $\phi\otimes\psi\in\kernel{\rho}$). Therefore the rank of $\rho^{T_1}$ has to be $4$ when $\kernel{\rho}$ intersects the Segre variety transversely. If not, we know from Lemmas and \ref{lemmaSR2} and \ref{lemmatangentSegre}  that there are two options: i) it is possible to write $\rho$ as $\rho'+\lambda\proj{\zeta\otimes\xi}{\zeta\otimes\xi}$, where $\lambda$ and $\rho'$ is a rank $3$ PPT state supported on a $2\times 3$ or smaller subspace of $\mathbbm{C}^3\otimes\mathbbm{C}^3$, with $\rank{\rho'}=3$ and $\rank{\left(\rho'\right)^{T_1}}=d-1$, ii) $\rho$ is supported on a $2\times 3$ or smaller subspace itself. Option ii) is excluded because of the assumption of $\rho$ supported on $\mathbbm{C}^3\otimes\mathbbm{C}^3$. Our aim in the following will be to show that $\rank{\rho'}=\rank{\left(\rho'\right)^{T_1}}$ unless $\rho'$ is supported on a $2\times 2$ subspace, which is precisely the second possibility we allow in the proposition. First, observe that if $\rho'$ is supported on a $2\times 3$ subspace, we can use an analogue of Lemma \ref{lemmaSR2}. Either we have $\rho'=\rho''+\lambda'\proj{\zeta'\otimes\xi'}{\zeta'\otimes\xi'}$ where $\lambda'>0$ and $\rho''$ is supported on a $2\times 2$, $1\times 3$ or $1\times 2$ subspace, $\rank{\rho''}=2$ and $\rank{\left(\rho''\right)^{T_1}}=\rank{\left(\rho'\right)^{T_1}}-1$, or $\kernel{\rho'}$ intersects the respective Segre variety $\Sigma_{1,2}$ transversely.  In the latter case, by Bezout's Theorem the $3$-dimensional kernel of $\rho'$ has precisely three product vectors in it.  Actually, we can repeat the argument preceeding Lemmas \ref{lemmanotwotwo} and \ref{lemmaquadruples} to conclude that the product vectors in $\kernel{\rho'}$ have to be locally equivalent to 
\begin{equation}\label{threeproductvectors}
\left[\begin{array}{ccc}
\phi_1&\phi_2&\phi_3\\
\hline
\psi_1&\psi_2&\psi_3
\end{array}
\right]=
\left[\begin{array}{ccc}
1&0&1\\
0&1&1\\
\hline
1&0&0\\
0&1&0\\
0&0&1
\end{array}
\right]
\end{equation}
Obviously, these vectors span the kernel. We see that there are, within the $2\times 3$ subspace, only three product vectors in $\range{\rho'}=\left(\kernel{\rho'}\right)^{\bot}$. They are locally equivalent to
\begin{equation}\label{anotherthree}
\left[\begin{array}{ccc}
\zeta_1&\zeta_2&\zeta_3\\
\hline
\xi_1&\xi_2&\xi_3
\end{array}
\right]=
\left[\begin{array}{ccc}
0&1&1\\
1&0&-1\\
\hline
1&0&0\\
0&1&0\\
0&0&1
\end{array}
\right].
\end{equation}
Since $\rho'$ is separable and of rank $3$, it must be locally equivalent to a convex sum of projections onto  the vectors $\zeta_i\otimes\xi_i$ in \eqref{anotherthree}, which implies that $\tilde\rho^{T_1}$ is an analogous sum of projections onto  $\zeta_i^{\ast}\otimes\xi_i$. But $\zeta_i^{\ast}\otimes\xi_i=\zeta_i\otimes\xi_i$ if the product vectors are as in \eqref{anotherthree}. Therefore  $\range{\rho'}=\rank{\left(\rho'\right)^{T_1}}$, which implies $\range{\rho}=\rank{\rho^{T_1}}$, as expected. This proves our assertion for $\rho'$ supported on a $2\times 3$ subspace with $\kernel{\rho'}$ that intersects the corresponding Segre variety $\Sigma_{1,2}$ transversely. For the other nontrivial cases, we can have $\rho''$ separable and of rank $2$, supported on a $2\times 2$ subspace. There is also the trivial case of $\rho''$ supported on a $1\times 2$ or $1\times 3$ subpace, in which the equality $\rank{\rho''}=\rank{\left(\rho''\right)^{T_1}}$ clearly holds, and it implies equality of ranks of $\rho$ and $\rho^{T_1}$.

In the case of $\rho''$ supported on a $2\times 2$ subspace, we can repeat the argument with transverse intersections. Either $\rho''$ can be reduced once again, in which case it turns out to be equal to $\lambda'''\proj{\zeta'''\otimes\xi'''}{\zeta'''\otimes\xi'''}+\lambda''\proj{\zeta''\otimes\xi''}{\zeta''\otimes\xi''}$ with $\lambda''>0$, $\lambda'''>0$ and $\zeta'''\otimes\xi'''$ not proportional to $\zeta''\otimes\xi''$, or $\kernel{\rho''}$ must intersect the respective Segre variety $\Sigma_{1,1}$ in a transverse way. The first possibility clearly gives us $\rank{\rho''}=2=\rank{\left(\rho''\right)^{T_1}}$. The latter implies, by Bezout's Theorem, that there are exactly two product vectors in $\kernel{\rho''}$. Similarly as for \eqref{threeproductvectors}, we can prove that the two product vectors must be locally equivalent to $e_1\otimes e_1$ and $e_2\otimes e_2$. Clearly, they span the kernel of $\rho''$ and there are only two product vectors, locally equivalent to $e_1\otimes e_2$ and $e_2\otimes e_1$, in $\range{\rho''}$. But $\rho''$ is separable and of rank $2$. Therefore it must be locally equivalent to a convex sum of projections onto these two vectors. Accordingly, $\left(\rho''\right)^{T_1}$ is locally equivalent to a sum of two projections onto product vectors, which are $e_1^{\ast}\otimes e_2$ and $e_2^{\ast}\otimes e_1$, actually equal to $e_1\otimes e_2$ and  $e_2\otimes e_1$. This clearly implies $\rank{\rho''}=\rank{\left(\rho''\right)^{T_1}}$ and the equality between the ranks of $\rho$ and $\rho^{T_1}$ clearly follows.
\end{proof}
\end{proposition}

\begin{remark}\label{remarkone}
The two propositions above explain why PPT states of ranks $\left(4,n\right)$, $n\neq 4$ should not be expected to appear in the upper part of \textrm{Table II} in \cite{LS2010n}. They do exist, but they are always separable and of a rather special form. 
\end{remark}

It is useful to formulate the following
\begin{corollary}\label{coredge}
All rank $4$ non-separable PPT states $\rho$ in $3\times 3$ systems are edge states.
\begin{proof}
If some non-separable $\rho$ of rank $4$ had a product vector $\phi\otimes\psi$ in its range, and the partial conjugated vector $\phi^{\ast}\otimes\psi$ was in the range of $\rho^{T_1}$, we could diminish the rank of $\rho$ or $\rho^{T_1}$ by subtracting $\lambda\proj{\phi\otimes\psi}{\phi\otimes\psi}$, where
\begin{equation}\label{defLambda} \lambda=\min\left\{\innerpr{\phi\otimes\psi}{\rho^{-1}\left(\phi\otimes\psi\right)}^{-1},\innerpr{\phi\otimes\psi}{\left(\rho^{T_1}\right)^{-1}\left(\phi\otimes\psi\right)}^{-1}\right\},
\end{equation}  
cf. \cite{LKHC2001}. In such case, $\rho$ could be written as $\rho=\rho'+\lambda\proj{\phi\otimes\psi}{\phi\otimes\psi}$ with $\rho'$ PPT and of rank $3$ or with $\rho^{T_1}$ of rank $3$. But this implies, by \cite{HLVC2000}, that $\rho'$ would have to be separable. This further implies separability of $\rho$, which is a contradiction. 
\end{proof}
\end{corollary}
At this point, we can easily prove that the vectors $\phi_i\otimes\psi_i$ in the kernel of a non-separable $\rho$ of rank $4$, chosen as in \eqref{formtwoindeptriples2} or \eqref{formtwoindeptriples3}, constitute a generalized Unextendible Product Basis. If there was a product vector $\phi\otimes\psi$ orthogonal to all of them, it would be an element of the range of $\rho$. From the proof of Proposition \ref{propequalranks} we know that the partially conjugated vectors $\phi_i^{\ast}\otimes\psi_i$ span the kernel of $\rho^{T_1}$. Since $\innerpr{\phi\otimes\psi}{\phi_i\otimes\psi_i}=0=\innerpr{\phi^{\ast}\otimes\psi}{\phi_i^{\ast}\otimes\psi_i}$ for all $i$, we see that $\phi^{\ast}\otimes\psi$ is in the range of $\rho^{T_1}$, $\left(\kernel\rho^{T_1}\right)^{\bot}$. Therefore we have a product vector $\phi\otimes\psi$ in the range of $\rho$ such that its partial conjugation is in the range $\rho^{T_1}$. In other words, $\rho$ is not an edge state. But this contradicts Corollary \ref{coredge} and therefore cannot happen. In this way, we have proved the following.
\begin{proposition}\label{propgUPB}
Let $\rho$ be a rank $4$ non-separable PPT state in a $3\times 3$ system. The six vectors in the kernel of $\rho$ constitute a generalized UPB. There is a subset of five of them that constitutes a minimal gUPB in the sense of Proposition \ref{propgUPBsuffnecessary}.
\begin{proof}
Most of the proof has already been explained above. We only need to comment on the fact that five of the product vectors constitute a minimal gUPB. It must be so because the five vectors we brought to the form \eqref{formtwoindeptriples2} or \eqref{formtwoindeptriples3} span the kernel of $\rho$, and  the orthogonal complement to the kernel has no product vector in it. Thus, the five vectors are a gUPB of $\kernel{\rho}$, which is minimal according to Proposition \ref{propgUPBsuffnecessary}, because $m+n-1=5$ for $m=n=3$. 
\end{proof}
\end{proposition}
By Proposition \ref{propgUPBsuffnecessary} we know that a minimal gUPB $\left\{\phi_i\otimes\psi_i\right\}_{i=1}^6$ has the property that all triples in $\left\{\phi_i\right\}_{i=1}^6$ and in $\left\{\psi_i\right\}_{i=1}^6$ are linearly independent. In such case, the forms \eqref{formtwoindeptriples2} and \eqref{formtwoindeptriples3} are locally equivalent, and we may choose to work with only one of them. In the sequel, we prefer to assume the form  \eqref{formtwoindeptriples2} of the product vectors, which is in agreement with the convention used in \cite{HHMS2011}. Our next step is to prove that the parameters $p$, $q$, $r$ and $s$ in \eqref{formtwoindeptriples2} must be real if the corresponding product vectors belong to the kernel of a rank $4$ PPT state in the $3\times 3$ case. This is not of much use here, but will prove to be important in Section \ref{secdetermination}.

We know from Lemma \ref{lemmageneralintersection} that there are exactly six product vectors in the kernel of $\rho$, while we have only five of them in \eqref{formtwoindeptriples2}, and we know that they span the kernel. Consequently, the sixth vector is a linear combination of the other five ones,
\begin{equation}\label{lincombsixth}
\phi_6\otimes\psi_6=\sum_{i=1}^5\lambda_i\phi_i\otimes\psi_i
\end{equation}
Note that explicit formulas for the sixth vector can be found in \cite[Section 5.2]{HHMS2011}. Interestingly, since $\phi_6\otimes\psi_6\in\kernel{\rho}$, we know from Lemma \ref{lemmaprodPPT} that $\phi_6^{\ast}\otimes\psi_6\in\kernel{\rho}$ is in the kernel of $\rho^{T_1}$. However, the vectors $\phi_i^{\ast}\otimes\psi_i$ with $i=1,2,\ldots,5$ are also there and moreover, since they are, up to local equivalence, of the form \eqref{formtwoindeptriples2} with $p$ and $q$ complex conjugated, we already know that they span $\kernel{\rho^{T_1}}$. Thus
\begin{equation}\label{lincombsixthconj}
\phi_6^{\ast}\otimes\psi_6=\sum_{i=1}^5\xi_i\phi_i^{\ast}\otimes\psi_i,
\end{equation}
where the coefficients $\xi_i$ are in principle not related to the $\lambda_i$'s in \eqref{lincombsixth}. However, we can already see at this point that it may be very difficult to simultaneously satisfy equations \eqref{lincombsixth} and \eqref{lincombsixthconj}, if we do not assume that $\phi_i=\phi_i^{\ast}$ for all $i$. In the latter case, one can obviously choose $\xi_i=\lambda_i$. Our aim in the following will be to show that $\xi_i=\lambda_i$ is the only possible choice. By projecting \eqref{lincombsixth} onto the first, the second and the third coordinate in the first subsystem, we get
\begin{eqnarray}\label{lambdaeqs1}
\lambda_1\psi_1+\lambda_4\psi_4+\lambda_5\psi_5&=&\phi_{16}\,\psi_6\\
\lambda_2\psi_2+\lambda_4\psi_4+p\lambda_5\psi_5&=&\phi_{26}\,\psi_6\label{lambdaeqs2}\\
\lambda_3\psi_3+\lambda_4\psi_4+q\lambda_5\psi_5&=&\phi_{36}\,\psi_6\label{lambdaeqs3}
\end{eqnarray}
where $\left\{\phi_{i6}\right\}_{i=1}^3$ are coordinates of $\phi_6$. Similarly, from \eqref{lincombsixthconj} we get
\begin{eqnarray}\label{xieqs1}
\xi_1\psi_1+\xi_4\psi_4+\xi_5\psi_5&=&\phi_{16}^{\ast}\,\psi_6\\
\xi_2\psi_2+\xi_4\psi_4+p^{\ast}\xi_5\psi_5&=&\phi_{26}^{\ast}\,\psi_6\label{xieqs2}\\
\xi_3\psi_3+\xi_4\psi_4+q^{\ast}\xi_5\psi_5&=&\phi_{36}^{\ast}\,\psi_6\label{xieqs3}
\end{eqnarray}
Let us note that the triples $\left\{\psi_1,\psi_4,\psi_5\right\}$, $\left\{\psi_2,\psi_4,\psi_5\right\}$, $\left\{\psi_3,\psi_4,\psi_5\right\}$ all consist of linearly independent vectors, according to Proposition \ref{propgUPB}. This implies that each of the formulas \eqref{lambdaeqs1}--\eqref{xieqs3} gives exactly one solution for the coefficients $\lambda_i$ or $\xi_i$ which it contains. For one of the consequences, all the coefficients $\psi_{i6}$ must be non-vanishing. Two of them cannot vanish, because $\phi_6$ proportional to any of $\phi_i$ with $i=1,2,3$ would contradict $\phi_6\otimes\psi_6\neq\phi_i\otimes\psi_i$ or Lemma \ref{lemma3x3twoproducts}. To see this, let us assume that one of them vanishes, e.g. $\phi_{36}=0$. In such case, equation \eqref{lambdaeqs3} implies $\lambda_3=\lambda_4=\lambda_5=0$, where we used the fact that $q\neq 0$. Hence \eqref{lambdaeqs1} and \eqref{lambdaeqs2} reduce to $\phi_{16}\psi_6=\lambda_1\psi_1$ and $\phi_{26}\psi_6=\lambda_2\psi_2$. But neither of these equalities can hold, since $\phi_{16}\neq 0$ and $\phi_{26}\neq 0$ and $\psi_6$ proportional to $\psi_1$ or $\psi_2$ contradicts Lemma~\ref{lemma3x3twoproducts}. Thus our assumption $\phi_{36}=0$ must have been false. By repeating the same argument for $\phi_{16}$ and $\phi_{26}$, we arrive at $\phi_{16}\phi_{26}\phi_{36}\neq 0$. Let us also notice that necessarily $\lambda_4\neq 0$ and $\xi_4\neq 0$. We cannot have, for example $\xi_1\psi_1+\xi_5\psi_5=\phi_{16}^{\ast}\psi_6$ and $\xi_2\psi_2+p^{\ast}\xi_5\psi_5=\phi_{26}^{\ast}\psi_6$ since the only vector in the intersection of $\linspan{\psi_1,\psi_5}$ and $\linspan{\psi_2,\psi_5}$ is $\psi_5$, and we know that $\psi_6\neq\psi_5$ by Lemma \ref{lemma3x3twoproducts} or simply by $\phi_6\otimes\psi_6\neq\phi_i\otimes\psi_i$. In a similar way, one obtains $\lambda_5\neq 0$ and $\xi_5\neq 0$. With such amount of knowledge, we can easily prove the expected result.
\begin{proposition}\label{proprealparams}
Let $\phi_i\otimes\psi_i$ for $i=1,2,\ldots,5$ be product vectors of the form \eqref{formtwoindeptriples2} in the kernel of a non-separable PPT state of rank four, acting on $\mathbbm{C}^3\otimes\mathbbm{C}^3$. The parameters $p$, $q$, $r$ and $s$ must necessarily be real.
\begin{proof}
By dividing \eqref{lambdaeqs1} by $\phi_{16}$ and \eqref{xieqs1} by  $\phi_{16}^{\ast}$, which is possible according to $\phi_{16}\neq 0$, we get
\begin{equation}\label{eqpsisixeq}
\frac{\lambda_1}{\phi_{16}}\psi_1+\frac{\lambda_4}{\phi_{16}}\psi_4+\frac{\lambda_5}{\phi_{16}}\psi_5=\psi_6=\frac{\xi_1}{\phi_{16}^{\ast}}\psi_1+\frac{\xi_4}{\phi_{16}^{\ast}}\psi_4+\frac{\xi_5}{\phi_{16}^{\ast}}\psi_5
\end{equation}
Since $\left\{\psi_1,\psi_4,\psi_5\right\}$ is a linearly independent triple, the above equality implies $\lambda_1/\phi_{16}=\xi_1/\phi_{16}^{\ast}$, $\lambda_4/\phi_{16}=\xi_4/\phi_{16}^{\ast}$ and $\lambda_5/\phi_{16}=\xi_5/\phi_{16}^{\ast}$. In a similar way, from \eqref{lambdaeqs2} and \eqref{xieqs2} we can get $\lambda_2/\phi_{26}=\xi_2/\phi_{26}^{\ast}$, $\lambda_4/\phi_{26}=\xi_4/\phi_{26}^{\ast}$ and $p\lambda_5/\phi_{26}=p^{\ast}\xi_5/\phi_{26}^{\ast}$, whereas \eqref{lambdaeqs3} and \eqref{xieqs3} give us $\lambda_2/\phi_{36}=\xi_2/\phi_{36}^{\ast}$, $\lambda_4/\phi_{36}=\xi_4/\phi_{36}^{\ast}$ and $q\lambda_5/\phi_{36}=q^{\ast}\xi_5/\phi_{36}^{\ast}$. From the equalities involving $\lambda_4$ and $\xi_4$, we get
\begin{equation}\label{xifourlambdafour}
\frac{\phi_{16}}{\phi_{16}^{\ast}}=\frac{\phi_{26}}{\phi_{26}^{\ast}}=\frac{\phi_{36}}{\phi_{36}^{\ast}}
\end{equation}
Together with $\lambda_5/\phi_{16}=\xi_5/\phi_{16}^{\ast}$, the above equations give us $\lambda_5/\phi_{26}=\xi_5/\phi_{26}^{\ast}$ and $\lambda_5/\phi_{36}=\xi_5/\phi_{36}^{\ast}$. But 
\begin{equation}\label{eqimplication}
\left(\frac{\lambda_5}{\phi_{26}}=\frac{\xi_5}{\phi_{26}^{\ast}}\,\land\,\frac{p\lambda_5}{\phi_{26}}=\frac{p^{\ast}\xi_5}{\phi_{26}^{\ast}}\right)\,\Rightarrow\,p=p^{\ast}
\end{equation}
In a similar way, from $\lambda_5/\phi_{36}=\xi_5/\phi_{36}^{\ast}$ and $q\lambda_5/\phi_{36}=q^{\ast}\xi_5/\phi_{36}^{\ast}$ we can get $q=q^{\ast}$.
\end{proof}
\end{proposition}

\subsection{An equivalence between generalized and orthonormal Unextendible Product Bases}\label{secequiv}

In the following, we discuss item $\left(4\right)$ of the list given in Section \ref{secoutline}. Let us start with a set of five vectors in $\mathbbm{C}^3$,
\begin{equation}\label{fivevectors}
\left[\begin{array}{ccccc}
\phi_1&\phi_2&\phi_3&\phi_4&\phi_5
\end{array}\right]=\left[\begin{array}{ccccc}
\phi_{11}&\phi_{12}&\phi_{13}&\phi_{14}&\phi_{15}\\
\phi_{21}&\phi_{22}&\phi_{23}&\phi_{24}&\phi_{25}\\
\phi_{31}&\phi_{32}&\phi_{33}&\phi_{34}&\phi_{35}
\end{array}\right],
\end{equation}
and assume that any three of them are linearly independent, as in Proposition \ref{propgUPB}. For the moment, we do not require the vectors in \eqref{fivevectors} to be equal to $\phi_1,\ldots,\phi_5$ in \eqref{formtwoindeptriples2}, but our ultimate goal is to apply the results we are going to obtain to \eqref{formtwoindeptriples2}. $\textnormal{PSL}\left(3,\mathbbm{C}\right)$ transformations of the above set correspond to the multiplication of the matrix in \eqref{fivevectors} from the left by an element of $\textnormal{SL}\left(3,\mathbbm{C}\right)$ and to the multiplication of the columns of \eqref{fivevectors} by arbitrary non-zero scalar factors. It is clear that we can $\textnormal{PSL}\left(3,\mathbbm{C}\right)$ transform \eqref{fivevectors} to the following form,
\begin{equation}\label{fivevectors2}
\left[\begin{array}{ccccc}
1&0&\phi'_{13}&\phi'_{14}&\phi'_{15}\\
0&1&\phi'_{23}&\phi'_{24}&\phi'_{25}\\
0&0&\phi'_{33}&\phi'_{34}&\phi'_{35}
\end{array}\right].
\end{equation}
By another $\textnormal{PSL}\left(3,\mathbbm{C}\right)$ transformation, we get
\begin{equation}\label{fivevectors3}\left[\begin{array}{ccc}1&0&0\\0&1&-\frac{\phi'_{23}}{\phi'_{33}}\\0&0&1\end{array}\right]\left[\begin{array}{ccccc}
1&0&\phi'_{13}&\phi'_{14}&\phi'_{15}\\
0&1&\phi'_{23}&\phi'_{24}&\phi'_{25}\\
0&0&\phi'_{33}&\phi'_{34}&\phi'_{35}
\end{array}\right]=
\left[\begin{array}{ccccc}
1&0&\phi'_{13}&\phi''_{14}&\phi''_{15}\\
0&1&0&\phi''_{24}&\phi''_{25}\\
0&0&\phi'_{33}&\phi''_{34}&\phi''_{35}
\end{array}\right].
\end{equation}
We should remark that the matrix we multiply with from the left is well-defined, since $\phi'_{33}\neq 0$ according to the assumption about linear independence of triples. Let us transform once again, in the following way, 
\begin{equation}\label{fivevectors4}\left[\begin{array}{ccc}1&0&-\frac{\phi''_{15}}{\phi''_{35}}\\0&1&0\\0&0&1\end{array}\right]\left[\begin{array}{ccccc}
1&0&\phi'_{13}&\phi''_{14}&\phi''_{15}\\
0&1&0&\phi''_{24}&\phi''_{25}\\
0&0&\phi'_{33}&\phi''_{34}&\phi''_{35}
\end{array}\right]=
\left[\begin{array}{ccccc}
1&0&\phi'_{13}&\phi'''_{14}&0\\
0&1&0&\phi'''_{24}&\phi''_{25}\\
0&0&\phi'_{33}&\phi'''_{34}&\phi''_{35}
\end{array}\right].
\end{equation}
This is again possible because $\phi'_{35}\neq 0$ according to our assumptions. 

In a similar way as before, we see that $\phi''_{24}\neq 0$ and $\phi''_{35}\neq 0$. If we multiply the fourth column by ${1}/{\phi'''_{24}}$ and the fifth by ${1}/{\phi''_{35}}$, the above transforms to
\begin{equation}\label{fivevectors5}
\left[\begin{array}{ccccc}
1&0&x&y&0\\
0&1&0&1&z\\
0&0&t&u&1
\end{array}\right],
\end{equation}
where we introduced the notation $x:=\phi'_{13}$, $t:=\phi'_{33}$, $y:={\phi'''_{14}}/{\phi''_{24}}$, $u:={\phi'''_{34}}/{\phi''_{24}}$, $z:={\phi''_{25}}/{\phi''_{35}}$. It is pretty straightforward to see that all the coefficients $x,y,z,t,u$ have to be different from zero according to the independent triples assumption.

Now, introduce the following invariants \cite{LS2010},
\begin{equation}\label{invariants1}
s_1=-\,\frac{\left|\begin{array}{ccc}\phi_{1}&\phi_{2}&\phi_{4}\end{array}\right|\cdot\left|\begin{array}{ccc}\phi_{1}&\phi_{3}&\phi_{5}\end{array}\right|}{\left|\begin{array}{ccc}\phi_{1}&\phi_{2}&\phi_{5}\end{array}\right|\cdot\left|\begin{array}{ccc}\phi_{1}&\phi_{3}&\phi_{4}\end{array}\right|},
\end{equation}
\begin{equation}\label{invariants2}
s_2=-\,\frac{\left|\begin{array}{ccc}\phi_{1}&\phi_{2}&\phi_{3}\end{array}\right|\cdot\left|\begin{array}{ccc}\phi_{2}&\phi_{4}&\phi_{5}\end{array}\right|}{\left|\begin{array}{ccc}\phi_{1}&\phi_{2}&\phi_{4}\end{array}\right|\cdot\left|\begin{array}{ccc}\phi_{2}&\phi_{3}&\phi_{5}\end{array}\right|}.
\end{equation}
The numbers $s_1$, $s_2$ are indeed invariant. They do not change under the family of $\textnormal{PSL}\left(3,\mathbbm{C}\right)$ transformations we were using in the consecutive steps \eqref{fivevectors}--\eqref{fivevectors5}. Thus we can substitute
\begin{equation}\label{subst}
\left[\begin{array}{ccccc}
\phi_{11}&\phi_{12}&\phi_{13}&\phi_{14}&\phi_{15}\\
\phi_{21}&\phi_{22}&\phi_{23}&\phi_{24}&\phi_{25}\\
\phi_{31}&\phi_{32}&\phi_{33}&\phi_{34}&\phi_{35}
\end{array}\right]\rightarrow\left[\begin{array}{ccccc}
1&0&x&y&0\\
0&1&0&1&z\\
0&0&t&u&1
\end{array}\right]
\end{equation}
in the above formulas for $s_1$ and $s_2$. In this way, we can easily calculate the values of the invariants,
\begin{equation}\label{valuesofinvariants}
s_1=-uz\quad\textrm{and}\quad s_2=-\frac{ty}{ux}.
\end{equation}
Now, impose the conditions $s_1>0$ and $s_2>0$. From the first one, we clearly get $u=-rz^{\ast}$, where $r$ is a positive real number. Thus, we have the vectors
\begin{equation}\label{fivevectors6}
\left[\begin{array}{ccccc}
1&0&x&y&0\\
0&1&0&1&z\\
0&0&t&-rz^{\ast}&1
\end{array}\right].
\end{equation}
Next, let us multiply from the left by a diagonal matrix $\textnormal{diag}\left(1,\sqrt{\sqrt{r'}},{1}/{\sqrt{\sqrt{r'}}}\right)$, as well as multiply the second column by ${1}/{\sqrt{\sqrt{r'}}}$, the fourth by ${1}/{\sqrt{\sqrt{r'}}}$ and the fifth by $\sqrt{\sqrt{r'}}$, where $r':=r{z^{\ast}}/{z}$ and $\sqrt{\zeta}$ stands for the square root of $\zeta\in\mathbbm{C}$ with the argument in $\left[0,\pi\right)$. Under such $\textnormal{PSL}\left(3,\mathbbm{C}\right)$ transformation \eqref{fivevectors6} changes to 
\begin{equation}\label{fivevectors7}
\left[\begin{array}{ccccc}
1&0&x'&y'&0\\
0&1&0&1&z'\\
0&0&t'&-z'&1
\end{array}\right],
\end{equation}
where $z'$ is real and positive, and all the other parameters $x',y',t'$ are non-zero. Moreover, the conditon $s_2>0$ transforms to
\begin{equation}\label{conds2}
s_2=-\frac{ty}{ux}=\frac{t'y'}{z'x'}>0\quad\Leftrightarrow\quad\frac{t'y'}{x'}>0,
\end{equation}
simply by formula \eqref{valuesofinvariants} and the invariance of $s_2$. The last equivalence holds by strict positivity of $z'$. In our next step, we we are going to multiply \eqref{fivevectors7} from the left by a diagonal matrix $\textnormal{diag}\left(\zeta_1,\zeta_2,\zeta_3\right)$, with $\zeta_1,\zeta_2,\zeta_3\in\mathbbm{C}$ and $\zeta_1\zeta_2\zeta_3=1$, and also multiply the consecutive columns, beginning with the first, by ${1}/{\zeta_1}$, ${1}/{\zeta_2}$, $\zeta_4$, $\zeta_5$ and $\zeta_6$, where $\zeta_4\zeta_5\zeta_6\neq 0$. Our aim is to choose the numbers $\zeta_1,\ldots,\zeta_6$ in such a way that $\eqref{fivevectors7}$ transforms to a set of vectors with orthogonality relations given by a pentagon graph (that is, any two consecutive ones are orthogonal, and these are the only orthogonality relations). We would like to have
\begin{equation}\label{fivevectors8}
\left[\begin{array}{ccccc}
1&0&a&b&0\\
0&1&0&1&a\\
0&0&b&-a&1
\end{array}\right],
\end{equation}
where $a=z'$ and $b$ is a positive real number in place of \eqref{fivevectors7}. Let us write the numbers $\zeta_j$ as $r_je^{i\alpha_j}$, where $r_j$ is a positive real number and $\alpha_j\in\mathbbm{R}$. In order to obtain \eqref{fivevectors8} with $a$ and $b$ real and positive, certain phase matching conditions have to be fulfilled. Let us consider them first. If $\alpha_{y'}$, $\alpha_{t'}$, $\alpha_{x'}$ are such that $y'=r_{y'}e^{i\alpha_{y'}}$, $t'=r_{t'}e^{i\alpha_{t'}}$ and $x'=r_{x'}e^{i\alpha_{x'}}$ with $r_{x'}$, $r_{t'}$ and $r_{x'}$ real and positive, complex phases match correctly if and only if the following set of equations hold
\begin{eqnarray}
\alpha_2+\alpha_5&=&0\quad\textnormal{ mod }2\pi\label{eq1}\\
\alpha_3+\alpha_6&=&0\quad\textnormal{ mod }2\pi\label{eq2}\\
\alpha_2+\alpha_6&=&0\quad\textnormal{ mod }2\pi\label{eq3}\\
\alpha_3+\alpha_5&=&0\quad\textnormal{ mod }2\pi\label{eq4}\\
\alpha_5+\alpha_1+\alpha_{y'}&=&0\quad\textnormal{ mod }2\pi\label{eq5}\\
\alpha_3+\alpha_4+\alpha_{t'}&=&0\quad\textnormal{ mod }2\pi\label{eq6}\\
\alpha_4+\alpha_1+\alpha_{x'}&=&0\quad\textnormal{ mod }2\pi\label{eq7}
\end{eqnarray}
The requirement that $\zeta_1\zeta_2\zeta_3=1$ adds a condition $\alpha_1+\alpha_2+\alpha_3=0\textnormal{ mod }2\pi$ to equations \eqref{eq1}--\eqref{eq2}. However, a substitution of the form
\begin{equation}\nonumber
\left(\alpha_1,\alpha_2,\alpha_3\right)\rightarrow\left(\alpha_1+\beta,\alpha_2+\beta,\alpha_3+\beta\right)\quad\left(\alpha_4,\alpha_5,\alpha_6\right)\rightarrow\left(\alpha_4-\beta,\alpha_5-\beta,\alpha_6-\beta\right)
\end{equation}
with an appropriately chosen $\beta$ can always bring $\alpha_1+\alpha_2+\alpha_3$ to zero and it has no effect on \eqref{eq1}--\eqref{eq7}. Therefore, as long as existence of solutions is in question, we may neglect the additional condition. It is easy to see that the relations \eqref{eq1}--\eqref{eq4} are fulfilled if and only if $\alpha_2=\alpha_3=-\alpha_5=-\alpha_6=-\alpha\textnormal{ mod }2\pi$ for some $\alpha\in\mathbbm{R}$. Thus the set of equations \eqref{eq1}--\eqref{eq7} are reduced to 
\begin{equation}\label{matrixeq}
\left[\begin{array}{ccc}1&1&0\\-1&0&1\\0&1&1\end{array}\right]\left[\begin{array}{c}\alpha\\\alpha_1\\\alpha_4\end{array}\right]=\left[\begin{array}{c}-\alpha_{y'}\\-\alpha_{t'}\\-\alpha_{x'}\end{array}\right]\textnormal{ mod }2\pi
\end{equation}
Interestingly, the $3\times 3$ matrix in equation \eqref{matrixeq} has rank $2$. A solution $\left(\alpha,\alpha_1,\alpha_4\right)$ exists if and only if 
\begin{equation}\label{phasematchingfinal}
\alpha_{y'}+\alpha_{t'}-\alpha_{x'}=0\textnormal{ mod }2\pi
\end{equation} 
But this is exactly the positivity condition \eqref{conds2} for the invariant $s_2$. Thus, if $s_2>0$ in addition to $s_1>0$, we can cancel the complex phases, as in \eqref{fivevectors8}. The only remaining thing to do  is to match the modules, which gives us the following set of equations,
\begin{equation}\label{eqsmodulematch}
r_2r_5=1,\quad r_3r_6=1,\quad r_2r_6=r_3r_5,\quad r_4r_1r_{x'}=a,\quad r_5r_1r_{y'}=r_3r_4r_{t'}.
\end{equation}
There is also an equation $r_1r_2r_3=1$, following from the requirement that $\zeta_1\zeta_2\zeta_3=1$. As we see, there are five equations in \eqref{eqsmodulematch}, and the variables $r_1,\ldots,r_6$ are six in number. Therefore, one can expect a solution to exist. It can easily be checked that the following, with $r\in\mathbbm{R}$, is a one-parameter family of solutions,
\begin{equation}\label{solutionmodulesmatch}
r_1=\sqrt{\frac{ar_{t'}}{r_{x'}r_{y'}}}r,\quad r_2=r,\quad r_3=r, \quad r_4=\sqrt{\frac{ar_{t'}}{r_{x'}r_{y'}}}\frac{1}{r}, \quad r_5=\frac{1}{r},\quad r_6=\frac{1}{r}.
\end{equation}
By choosing $r=1/\sqrt[6]{{ar_{t'}}/{r_{x'}r_{y'}}}$ we can satisfy the additional condition $r_1r_2r_3=1$. Thus we have proved that the positivity of the invariants $s_1$, $s_2$ guarantees that the family of five vectors \eqref{fivevectors} can be $\textnormal{PSL}\left(3,\mathbbm{C}\right)$ transformed into \eqref{fivevectors8} without permuting them. Obviously, a converse statement is also true, since the values of $s_1$ and $s_2$ calculated from \eqref{fivevectors8} are $a^2$ and $b^2/a^2$, respectively. In this way we arrive at the following
\begin{proposition}\label{pentagon}
A set of five vectors $\left\{\alpha_i\right\}_{i=1}^5\subset\mathbbm{C}^3$ with the property that any triple of them is linearly independent, can be $\textnormal{PSL}\left(3,\mathbbm{C}\right)$ transformed, without permuting them, to the form \eqref{fivevectors8} with $a$ and $b$ real and positive, if and only if the invariants $s_1$ and $s_2$, defined in \eqref{invariants1}, are positive.
\end{proposition}
Let us note that any set of five vectors $\left\{v_1,\ldots,v_5\right\}\subset\mathbbm{C}^3$ with orthogonality relations $\innerpr{v_i}{v_{\left(i+1\right)\textnormal{ mod }5}}=0$ can be transformed by $\PSLt$ transformations to the form \eqref{fivevectors8}. For sure they can be transformed to
\begin{equation}\label{fivevectors9}
\left[\begin{array}{ccccc}v_1&v_2&v_3&v_4&v_5\end{array}\right]=\left[\begin{array}{ccccc}
1&0&x&y^{\ast}&0\\
0&1&0&1&x\\
0&0&y&-x^{\ast}&1
\end{array}\right],
\end{equation}
with $x$ and $y$ complex. But since $s_1=\left|x\right|^2>0$ and $s_2=\left|y/x\right|^2>0$ in the above case, the argument following equation \eqref{fivevectors7} tells us that a $\PSLt$ transformation brings \eqref{fivevectors9} to the form \eqref{fivevectors8}. As a consequence, Proposition \ref{pentagon} is a necessary and sufficient criterion for a set of five vectors $\phi_1,\ldots,\phi_5$ to be $\SLt$ equivalent, without permuting them, to a set of vectors  $v_1,\ldots,v_5$ with orthogonality relations $\innerpr{v_i}{v_{i\textnormal{ mod }5+1}}=0$.

From \cite{DiVicenzo04} we know that orthogonal UPBs in the $3\times 3$ case always have five elements, and they are, up to permutations, precisely the sets of product vectors $\left\{v_i\otimes w_i\right\}_{i=1}^5$ with orthogonality relations $\innerpr{v_i}{v_{i\textnormal{ mod }5+1}}=0$ and $\innerpr{w_j}{w_{\left(j+1\right)\textnormal{ mod }5+1}}=0$. Consider the question whether an arbitrary set of five vectors $\left\{\phi_i\otimes\psi_i\right\}_{i=1}^5\subset\mathbbm{C}^3\otimes\mathbbm{C}^3$ with linearly independent triples can be brought by $\PSLtt$ transformations to such $\left\{v_i\otimes w_i\right\}_{i=1}^5$, without permuting the vectors. In other words, what are the necessary and sufficient conditions for $\phi_i\otimes\psi_i$'s to be convertible into $v_i\otimes w_i$'s with the orthogonality conditions given above. By using Proposition \ref{pentagon}, we can already deal with the question about $\phi_i$'s being convertible into $v_i$'s. Namely, an $\PSLt$ transformation on the first subsystem can bring the vectors $\left\{\phi_i\right\}_{i=1}^5$, without permuting them, to $\left\{v_i\right\}_{i=1}^5$ with $\innerpr{v_i}{v_{i\textnormal{ mod }5+1}}=0$ if and only if the corresponding values of the invariants $s_1$ and $s_2$ are positive. We are only missing a similar criterion for $\psi_i$'s and $w_i$'s. However, it is not difficult to check that a permutation $\sigma=\left(\begin{array}{ccccc}1&2&3&4&5\\1&3&5&2&4\end{array}\right)$ brings any $\left\{w_i\right\}_{i=1}^5$ with $\innerpr{w_j}{w_{\left(j+1\right)\textnormal{ mod }5+1}}=0$ to $\left\{w'_i\right\}_{i=1}^5=\left\{w_{\sigma\left(i\right)}\right\}_{i=1}^5$ with $\innerpr{w'_i}{w'_{i\textnormal{ mod }5+1}}=0$. Therefore, it is sufficient to calculate the invariants \eqref{invariants1} and \eqref{invariants2} corresponding to the permuted vectors $\psi'_i:=\psi_{\sigma\left(i\right)}$ and check their positivity in order to tell whether the vectors $\psi_i$ are convertible into some $\left\{w_i\right\}_{i=1}^5$ with the desired orthogonality relations. Following the definitions \eqref{invariants1} and \eqref{invariants2}, let us introduce
\begin{multline}\label{invariant3}
s_3=
-\,\frac{\left|\begin{array}{ccc}\psi_{1}&\psi_{3}&\psi_{2}\end{array}\right|\cdot\left|\begin{array}{ccc}\psi_{1}&\psi_{5}&\psi_{4}\end{array}\right|}{\left|\begin{array}{ccc}\psi_{1}&\psi_{3}&\psi_{4}\end{array}\right|\cdot\left|\begin{array}{ccc}\psi_{1}&\psi_{5}&\psi_{2}\end{array}\right|}=\\
-\,\frac{\left|\begin{array}{ccc}\psi_{\sigma\left(1\right)}&\psi_{\sigma\left(2\right)}&\psi_{\sigma\left(4\right)}\end{array}\right|\cdot\left|\begin{array}{ccc}\psi_{\sigma\left(1\right)}&\psi_{\sigma\left(3\right)}&\psi_{\sigma\left(5\right)}\end{array}\right|}{\left|\begin{array}{ccc}\psi_{\sigma\left(1\right)}&\psi_{\sigma\left(2\right)}&\psi_{\sigma\left(5\right)}\end{array}\right|\cdot\left|\begin{array}{ccc}\psi_{\sigma\left(1\right)}&\psi_{\sigma\left(3\right)}&\psi_{\sigma\left(4\right)}\end{array}\right|}
\end{multline}
and
\begin{multline}\label{invariant4}
s_4=-\,\frac{\left|\begin{array}{ccc}\psi_{1}&\psi_{3}&\psi_{5}\end{array}\right|\cdot\left|\begin{array}{ccc}\psi_{3}&\psi_{2}&\psi_{4}\end{array}\right|}{\left|\begin{array}{ccc}\psi_{1}&\psi_{3}&\psi_{2}\end{array}\right|\cdot\left|\begin{array}{ccc}\psi_{3}&\psi_{5}&\psi_{4}\end{array}\right|}=\\
-\,\frac{\left|\begin{array}{ccc}\psi_{\sigma\left(1\right)}&\psi_{\sigma\left(2\right)}&\psi_{\sigma\left(3\right)}\end{array}\right|\cdot\left|\begin{array}{ccc}\psi_{\sigma\left(2\right)}&\psi_{\sigma\left(4\right)}&\psi_{\sigma\left(5\right)}\end{array}\right|}{\left|\begin{array}{ccc}\psi_{\sigma\left(1\right)}&\psi_{\sigma\left(2\right)}&\psi_{\sigma\left(4\right)}\end{array}\right|\cdot\left|\begin{array}{ccc}\psi_{\sigma\left(2\right)}&\psi_{\sigma\left(3\right)}&\psi_{\sigma\left(5\right)}\end{array}\right|},
\end{multline}
in accordance with \cite{LS2010}. From the discussion above it follows that arbitrary five vectors $\psi_1,\ldots,\psi_5$ in $\mathbbm{C}^3$ can be transformed, without permuting them, to $\left\{w_i\right\}_{i=1}^5$ with orthogonality relations $\innerpr{w_j}{w_{\left(j+1\right)\textnormal{ mod }5+1}}=0$ if and only if the above invariants $s_3$ and $s_4$ are positive. Together with the previously obtained convertibility result between $\phi_1,\ldots,\phi_5$ and $v_1,\ldots,v_5$, the last result gives us the following
\begin{proposition}\label{propinvariantspositiveortho}
A set of product vectors $\left\{\phi_i\otimes\psi_i\right\}_{i=1}^5\subset\mathbbm{C}^3\otimes\mathbbm{C}^3$ can be $\PSLtt$ transformed to an orthogonal UPB $\left\{v_i\otimes w_i\right\}_{i=1}^5$ with orthogonality relations $\innerpr{v_i}{v_{i\textnormal{ mod }5+1}}=0$ and $\innerpr{w_j}{w_{\left(j+1\right)\textnormal{ mod }5+1}}=0$, without permuting the $\phi_i\otimes\psi_i$'s, if and only if the invariants $s_1$, $s_2$, $s_3$ and $s_4$, defined in \eqref{invariants1}, \eqref{invariants2}, \eqref{invariant3} and \eqref{invariant4}, are positive.
\begin{proof}
Most of the proof has already been included above. Let $\left\{v_i\otimes w_i\right\}_{i=1}^5$ denote an orthogonal UPB  with the orthogonality relations $\innerpr{v_i}{v_{i\textnormal{ mod }5+1}}=0$ and $\innerpr{w_j}{w_{\left(j+1\right)\textnormal{ mod }5+1}}=0$ for all $i,j\in\left\{1,2,3,4,5\right\}$. The possibility to convert
\begin{equation}\label{posconvert}
\left[\begin{array}{ccccc}\phi_1&\phi_2&\phi_3&\phi_4&\phi_5\\
\hline
\psi_1&\psi_2&\psi_3&\psi_4&\psi_5
\end{array}
\right]\rightarrow\left[\begin{array}{ccccc}
v_1&v_2&v_3&v_4&v_5\\
\hline
w_1&w_2&w_3&w_4&w_5
\end{array}
\right]
\end{equation}
by $\PSLtt$ transformations, or by local equivalence in our usual terms, is the same as the possibility to separately convert $\left\{\phi_i\right\}_{i=1}^5$ into $\left\{v_i\right\}_{i=1}^5$ and $\left\{\psi_j\right\}_{j=1}^5$ into $\left\{w_j\right\}_{j=1}^5$ by some $\PSLt$ transformations. However, we know that the first conversion is possible if and only if $s_1$ and $s_2$ are positive, while the second needs positivity of $s_3$ and $s_4$. Altogether, positivity of all the invariants $s_i$, $i=1,2,3,4$ is a necessary and sufficient criterion for the transformation \eqref{posconvert} to be possible.  
\end{proof}
\end{proposition}
In the context of product vectors in the kernel of a PPT state, as well as elements of an orthogonal UPB, permutations are obviously possible. Therefore we would like to have a version of Proposition \ref{propinvariantspositiveortho} with no restriction on the ordering of the vectors $\left\{\phi_i\otimes\psi_i\right\}_{i=1}^5$. 
\begin{proposition}\label{propinvariantspositiveortho2}
A set of product vectors $\left\{\phi_i\otimes\psi_i\right\}_{i=1}^5\subset\mathbbm{C}^3\otimes\mathbbm{C}^3$ can be $\PSLtt$ transformed to an orthogonal UPB, if and only if for some permutation $\kappa$ the invariants $s_1$, $s_2$, $s_3$ and $s_4$, calculated with the permuted vectors $\phi_{\kappa\left(i\right)}$ and $\psi_{\kappa\left(i\right)}$ substituted for $\phi_i$ and $\psi_i$, respectively, are all positive.
\begin{proof}
Immediate given the fact \cite{DiVicenzo04} that an orthogonal UPB in a $3\times 3$ system can always be brought by a permutation to a $\left\{v_i\otimes w_i\right\}_{i=1}^5$ with the orthogonality relations as in Proposition \ref{propinvariantspositiveortho}.
\end{proof}
\end{proposition} 
Let us also note that, in accordance with \cite{HHMS2011}, not every single permutation of the five product vectors needs to be considered if we want to check whether they can be transformed into an orthogonal UPB or not.
\begin{remark}\label{remarkpermutation}
Only $12$ permutations, given in Appendix \ref{apppermutations}, have to be checked in Proposition \ref{propinvariantspositiveortho2} in order to obtain a decisive answer.
\begin{proof}
An explanation is included in \cite{LS2010} and \cite{HHMS2011}, but we repeat it quickly here for completness. Let us denote by $S_5$ the symmetric group of $\left\{1,2,\ldots,5\right\}$. The permutations given in Appendix \ref{apppermutations} are representatives of equivalence classes in $S_5$ of the regular pentagon subgroup $G$, generated by the cycle $\left(1\,2\,3\,4\,5\right)$ and the inversion $\left(\begin{array}{ccccc}1&2&3&4&5\\5&4&3&2&1\end{array}\right)$. The regular pentagon symmetry subgroup has the expected property that it does not change signs of $s_1$, $s_2$, $s_3$ and $s_4$, just as it does not change orthogonality relations between the vectors $\left\{v_i\right\}_{i=1}^5$ and $\left\{w_j\right\}_{j=1}^5$. Therefore, we may divide $S_5$ by $G$ when we check positivity of the invariants in Proposition \ref{propinvariantspositiveortho2}. The number of invariance classes is $12$ because $\#S_5=5!=120$ and $\#G=10$.
\end{proof}
\end{remark}

\subsection{Determination of a PPT state by product vectors in its kernel}\label{secdetermination}
In the last part of the proof, we recall a number of surprising facts that were earlier reported in \cite[Section 5]{HHMS2011} without a complete explanation. Here we fill in that little gap, and we collect a sufficient amount of knowledge to quickly obtain our main result.

Note that, given a set of product vectors in $\kernel{\rho}$, the conditions in Lemma \ref{lemmaprodPPT} are a set of linear equations for $\rho$. An idea, earlier presented in \cite{HHMS2011}, is to try to solve these equations assuming a specific form of the product vectors, namely \eqref{formtwoindeptriples2}. Let us repeat formula \eqref{formtwoindeptriples2} here for the convenience of the reader.
\begin{equation}\label{formtwoindeptriples4}
\left[\begin{array}{ccccc}\phi_1&\phi_2&\phi_3&\phi_4&\phi_5\\
\hline
\psi_1&\psi_2&\psi_3&\psi_4&\psi_5
\end{array}
\right]=\left[\begin{array}{ccccc}
1&0&0&1&1\\
0&1&0&1&p\\
0&0&1&1&q\\
\hline
1&0&0&1&1\\
0&1&0&1&r\\
0&0&1&1&s
\end{array}
\right],
\end{equation}
We actually know from Proposition \ref{propgUPB} that there always exists a local $\SLtt$ transformation $A\otimes B$ that brings five vectors in the kernel of a non-separable PPT state of rank $4$, possibly multiplied by some scalar factors, into the form \eqref{formtwoindeptriples4} with \emph{all triples linearly independent}. Moreover, Proposition \ref{proprealparams} tells us that the parameters $p$, $q$, $r$ and $s$ are necessarily real numbers. By solving the linear conditions on a PPT state following from Lemma \ref{lemmaprodPPT} with  $\phi_i\otimes\psi_i$, $i=1,2,\ldots,5$ as in \eqref{formtwoindeptriples4} substituted for $\phi\otimes\psi$, we will actually be solving a set of contraints on $\left(A^{-1}\otimes B^{-1}\right)^{\ast}\rho\left(A^{-1}\otimes B^{-1}\right)$. However, according to the discussion in Section \ref{secequivalence}, such local transformations are irrelevant to all the questions considered in this paper. Therefore we may simply assume that a PPT state $\rho$ in question has the product vectors \eqref{formtwoindeptriples4} in its kernel and check the consequences. As previously reported by the authors of \cite{HHMS2011}, the conditions $\innerpr{\phi_i\otimes\psi_j}{\rho\left(\phi_k\otimes\psi_i\right)}=0$ for $i,j,k\in\left\{1,2,3\right\}$ together with $\rho\left(\phi_4\otimes\psi_4\right)=0$ and $\innerpr{\phi_1\otimes\psi_4}{\rho\left(\phi_4\otimes\psi_2\right)}=0$ imply the following form of $\rho$,
\begin{equation}
\rho=\left[\begin{array}{ccc|ccc|ccc}
0&0&0&0&0&0&0&0&0\\
0&a_1&b_1&0&0&0&0&b_2&0\\
0&b_1&a_2&0&0&b_3&0&0&0\\
\hline
0&0&0&a_3&0&b_4&b_5&0&0\\
0&0&0&0&0&0&0&0&0\\
0&0&b_3&b_4&0&a_4&0&0&0\\
\hline
0&0&0&b_5&0&0&a_5&b_6&0\\
0&b_2&0&0&0&0&b_6&a_6&0\\
0&0&0&0&0&0&0&0&0
\end{array}\right]
\label{eqtabelaab}
\end{equation}
with $a_i$ and $b_j$ \emph{real} for all $i,j\in\left\{1,2,\ldots,6\right\}$ and such that
\begin{eqnarray}\label{eqslinearab1}
a_1+b_1+b_2=0,&b_1+a_2+b_3=0,&a_3+b_4+a_4=0\\
b_3+b_4+a_4=0,&b_5+a_5+b_6=0,&b_2+b_6+a_6=0\label{eqslinearab2}\\
&a_1+b_1+b_2=0&\label{eqslinearab3}
\end{eqnarray}
Derivation of \eqref{eqtabelaab} and the equations \eqref{eqslinearab1}--\eqref{eqslinearab3} is left as a simple exercise for the reader. It may be useful to consult Section 5.4 of \cite{HHMS2011} in order to solve it.

We still have not used the condition $\rho\left(\phi_5\otimes\psi_5\right)=0$, which gives us additional six linear equations on $a_1,\ldots,a_6$ and $b_1,\ldots,b_6$.
\begin{eqnarray}\label{eqlinearab4}
-r\left(b_1+ b_2\right)+q r b_2 + s b_1 = 0,& r b_1 - s\left(b_1+b_3\right) +p s b_3 = 0\\
-p\left(b_4 + b_5\right)+ q b_5  +p s b_4  = 0,&  p b_4  + s b_3  - p s\left(b_3 + b_4\right) = 0\label{eqlinearab5}\\ 
 p b_5  - q\left(b_5 + b_6\right)  + q r b_6  = 0,& q b_6  + r b_2 - q r\left(b_2 + b_6\right)  = 0\label{eqlinearab6}
\end{eqnarray}
Under the assumption of $\left\{\phi_i\otimes\psi_i\right\}_{i=1}^5$ of the form \eqref{formtwoindeptriples4} being a gUPB, there exists, up to scaling by arbitrary real factors, exactly one solution to the equations \eqref{eqslinearab1}--\eqref{eqlinearab6}. We know from Proposition \ref{propgUPB} that the assumption is true for vectors $\phi_i\otimes\psi_i$ in the kernel of a non-separable rank $4$ PPT state in $3\times 3$ systems. It is most important for us that there exist, up to scaling by arbitrary \emph{positive} factors, exactly two solutions
\begin{equation}\label{eqsolutionpm}\scalebox{0.85}{
$\pm\left[
\begin{array}{ccccccccc}
 0 & 0 & 0 & 0 & 0 & 0 & 0 & 0 & 0 \\
 0 & \frac{q r-s}{r\left(q-1\right)} & 1 & 0 & 0 & 0 & 0 & \frac{r-s}{r\left(1-q\right) } & 0 \\
 0 & 1 & \frac{r-p s}{s\left(1-p\right) } & 0 & 0 & \frac{r-s}{s\left(p-1\right)} & 0 & 0 & 0 \\
 0 & 0 & 0 & \frac{\left(r-s\right) \left(p s-q\right)}{p \left(p-q\right) \left(s-1\right)} & 0 & \frac{r-s}{p \left(1-s\right)} & \frac{r-s}{q-p} & 0 & 0 \\
 0 & 0 & 0 & 0 & 0 & 0 & 0 & 0 & 0 \\
 0 & 0 & \frac{r-s}{s\left(p-1\right)} & \frac{r-s}{p \left(1-s\right)} & 0 & \frac{\left(p-s\right)\left(r-s\right)}{p\left(p-1\right)  s\left(s-1\right) } & 0 & 0 & 0 \\
 0 & 0 & 0 & \frac{r-s}{q-p} & 0 & 0 & \frac{\left(q r-p\right)\left(r-s\right)}{q \left(1-q\right) \left(r-1\right)} & \frac{r-s}{q\left(q-1\right)} & 0 \\
 0 & \frac{r-s}{r\left(1-q\right) } & 0 & 0 & 0 & 0 & \frac{r-s}{q\left(q-1\right)} & \frac{\left(q-r\right)\left(r-s\right)}{q\left(1-q\right) r \left(r-1\right) } & 0 \\
 0 & 0 & 0 & 0 & 0 & 0 & 0 & 0 & 0
\end{array}
\right]$
}
\end{equation}
The above matrix is well-defined since all the numbers $p$, $q$, $r$, $s$, $p-1$, $q-1$, $r-1$, $s-1$, $p-q$ and $r-s$ are nonzero as a consequence of all triples of vectors in \eqref{formtwoindeptriples4} being linearly independent.

Note that, for both choices of sign, \eqref{eqsolutionpm} is a symmetric matrix. Moreover, it is \emph{symmetric with respect to partial transpose}. Therefore $\rho$ is PPT iff it is positive definite. A necessary condition for \eqref{eqsolutionpm} to be positive definite is that all the nonzero elements on its diagonal, as well as all nontrivial $2\times 2$ minors of the form $\left|\begin{array}{cc}\rho_{ii}&\rho_{ij}\\\rho_{ji}&\rho_{jj}\end{array}\right|$ are positive. Altogether, we have six nonzero elements on the diagonal
\begin{multline}\label{eqelondiag}
\pm\left\{\frac{q r - s}{r\left( q-1\right)},-\frac{r - p s}{s\left(p - 
 1\right)}, \frac{\left(r - s\right)\left(p s-q\right)}{p \left(p - q\right)\left( s-1\right)},\frac{\left(p - s\right)\left(r - 
   s\right)}{p \left(p-1\right)s\left(s-1\right)},\right.\\\left.  -\frac{\left( q r-p\right) (r - s)}{q\left (p-q\right)\left(r-1\right)}, \frac{\left(r-q\right)\left(r - s\right)}{q\left(q-1\right)r\left(r-1\right)}\right\}
\end{multline}
and six nontrivial minors
\begin{multline}\label{eqminors}
\left\{-\frac{\left(r - s\right)\left(q r - p s\right)}{r\left( p-1\right)s\left(q-1\right)}, -\frac{\left(q - s\right)\left(r - 
   s\right)}{q\left( q-1\right) r\left(r-1\right) }, \frac{\left(p - r\right)\left(r - s\right)}{p\left( p-1\right) s\left( 
   s-1\right)},\right.\\
   \left.\frac{\left(q - s\right) \left(r - s\right)^2}{p\left(p-1\right) \left(p - q\right)s\left(s-1\right) }, \frac{\left(r - 
   s\right)^2 \left(q r - p s\right)}{p\left(p - q\right) q\left(r-1\right)\left( 
   s-1\right)}, -\frac{\left(p - r\right)\left(r - s\right)^2}{\left(p - q\right)q\left(q-1\right) r \left(r-1\right) }\right\}
\end{multline}
The $\pm$ sign in \eqref{eqelondiag} corresponds to the choice we make in \eqref{eqsolutionpm}. We see that all the expressions in \eqref{eqminors} and \eqref{eqelondiag} are quotients and products of the following nineteen numbers
\begin{eqnarray}
&p,\,q,\,r,\,s,\,p-1,\,q-1,\,r-1,\,s-1,\,p-q,\,r-s,\label{numerkilista1}\\
&p-r,\,q-s,\,p-s,\,r-q,\,ps-q,\,qr-p,\,r-ps,\,qr-s,\,qr-ps\label{numerkilista2}
\end{eqnarray}
Concerning the list \eqref{numerkilista1}, we already know that all its elements have to be nonzero. This follows from the condition of $\left\{\phi_i\otimes\psi_i\right\}_{i=1}^5$ being a gUPB. It turns out that the same holds for the elements of \eqref{numerkilista2}. The number $qr-ps$  must be nonzero, because otherwise the vector 
\begin{equation}
\phi_5\otimes\psi_5-qr\phi_4\otimes\psi_4+q\left(s-r\right)\phi_3\otimes\psi_3-r\left(p-q\right)\phi_2\otimes\psi_2
\end{equation}
would be of the form $\phi_1\otimes\psi'+\phi'\otimes\psi_1$, thus contradicting Lemma \ref{lemmaSR2} and Corollary \ref{coredge}. In a similar way, one can show that $p-r\neq 0$ and $q-s\neq 0$. Let us now assume that $ps-q=0$. In such case, we have the following submatrix in \eqref{eqsolutionpm}
\begin{equation}\label{submatrix}
\pm\left[\begin{array}{cc}\frac{\left(r-s\right)\left(ps-q\right)}{p\left(p-q\right)\left(s-1\right)}&\frac{r-s}{p\left(1-s\right)}\\
\frac{r-s}{p\left(1-s\right)}&\frac{\left(p-s\right)\left(r-s\right)}{p\left(p-1\right)s\left(s-1\right)}
\end{array}\right]=\pm\left[\begin{array}{cc}0&\frac{r-s}{p\left(1-s\right)}\\
\frac{r-s}{p\left(1-s\right)}&\frac{\left(p-s\right)\left(r-s\right)}{p\left(p-1\right)s\left(s-1\right)}
\end{array}\right].
\end{equation}
In order for \eqref{submatrix} to be positive definite for some choice of the sign $\pm$, we need to have $r-s=0$, which we know is impossible. Thus we have proved that $ps-q\neq 0$ for $\rho$ positive definite. Finally, the fact that $qr-p$, $r-ps$ and $qr-s$ must also be nonvanishing for $\rho$ positive definite follows by a suitable modification of the above argument. Different submatrices need to be chosen, but otherwise the proof is identical. 

Our task in the following will be to relate positivity of all the numbers in \eqref{eqelondiag} and \eqref{eqminors} to the fact that all the invariants $s_1,\ldots,s_4$, given in Section \ref{secequiv}, are positive, possibly after we suitably permute the vectors $\phi_i\otimes\psi_i$. Note that we already know that only the $12$ permutations listed in Appendix \ref{apppermutations} need to be considered. An explanation is included in the proof related to Remark \ref{remarkpermutation}. Not to much surprise, the formulas for the invariants $s_1,\ldots,s_4$ for permuted vectors of the form \eqref{formtwoindeptriples4} are always expressed as products and quotients including only the numbers listed in \eqref{numerkilista1}. Explicit formulas can be found in Table \ref{tabinvariants}. To explain the notation we used in the table, it is sufficient to say, for example, that by using $\sigma_6$ from Appendix \ref{apppermutations} to permute the product vectors \eqref{formtwoindeptriples4}, we obtain $s_1=-p$, $s_2={\left(1-q\right)}/{q}$, $s_3=r-1$ and $s_4={r}/{\left(s-r\right)}$ as the expressions for the invariants. 
\begin{table}
\begin{tabular}{|l|l|l|l|}
\hline
$\sigma_1:$&$-\frac{p}{q}, q-1, \frac{r-s}{s}, \frac{r}{1-r}$&$\sigma_2:$&$-\frac{q}{p}, 
  p-1, \frac{s-r}{r}, \frac{s}{1-s}$\\
$\sigma_3:$&$-\frac{1}{q}, \frac{q-p}{p}, \frac{1-s}{s}, \frac{1}{r-1}$&$\sigma_4:$&$-q, \frac{1-p}{p},s-1,\frac{s}{r-s}$\\
$\sigma_5:$&$-\frac{1}{p}, \frac{p-q}{q}, \frac{1-r}{r}, \frac{1}{s-1}$&$\sigma_6:$&$-p,\frac{1-q}{q}, 
  r-1, \frac{r}{s-r}$\\
$\sigma_7:$&$\frac{p-q}{q}, \frac{1}{q-1}, -\frac{r}{s}, \frac{s-r}{r-1}$&$\sigma_8:$&$\frac{q}{p-q}, \frac{1-p}{q-1}, \frac{r}{s-r}, -s$\\
$\sigma_9:$&$-\frac{q-1}{q}, \frac{p}{q-p}, -\frac{1}{s}, \frac{1-s}{r-1}$&$\sigma_{10}:$&$\frac{q}{1-q}, \frac{p-1}{q-p}, \frac{1}{s-1}, -\frac{s}{r}$\\
$\sigma_{11}:$&$\frac{q-p}{p}, \frac{1}{p-1}, -\frac{s}{r}, \frac{r-s}{s-1}$&$\sigma_{12}:$&$\frac{p}{q-p}, \frac{1-q}{p-1}, \frac{s}{r-s}, -r$\\
\hline
\end{tabular}
\vskip 2 pt
\caption{Formulas for the invariants $s_1,\ldots,s_4$, calculated for vectors of the form \eqref{formtwoindeptriples4} permuted by the $12$ inequivalent permutations of Appendix \ref{apppermutations}.}
\label{tabinvariants}
\end{table}

It turns out that the values of $s_1,\ldots,s_4$ corresponding to one of the permutations $\sigma_i$ have to be all positive to assure that $\rho$, given in \eqref{eqsolutionpm}, is a positive matrix for some choice of the sign $\pm$. Our computer-aided proof of this fact consisted in simply checking all admissible sign choices for the numbers listed in \eqref{numerkilista1} and \eqref{numerkilista2}. We already know that neither of those numbers can be zero, and thus it seems that we have $2^{19}$ cases to check. However, some further constraints apply, which reduce this number considerably. First of all, the requirement that $\pm{\left(p - s\right)\left(r - 
   s\right)}/{\left(p \left(p-1\right)s\left(s-1\right)\right)}$ of the list \eqref{eqelondiag} and a very similar element ${\left(p - r\right)\left(r - s\right)}/{\left(p\left( p-1\right) s\left( 
   s-1\right)\right)}$ of \eqref{eqminors} have the same sign implies that $p-r=\pm\left(p-s\right)$, with the $\pm$ sign depending on the choice we made in \eqref{eqsolutionpm}. Along the same lines, by comparing the last element of \eqref{eqelondiag} with the second element of \eqref{eqminors}, one can prove that $r-q=\mp\left(q-s\right)$. More importantly, the signs of the numbers listed in \eqref{numerkilista1} and \eqref{numerkilista2} are not all independent. Various relations have to hold between them. For example, $p-1>0$ clearly implies $p>0$, and we cannot have a plus sign for $p-1$ and a minus sign for $p$. More sophisticated relations like
\begin{equation}\label{exemplaryrelation}
\left(qr-s>0\land p-1>0\land s<0\right)\Rightarrow qr-ps>0
\end{equation}
have to hold as well. We provide a more or less exhaustive list, consisting of 76 elements, in Appendix \ref{appconstraints}. While some further relations could still possibly exist, the use of those listed in the appendix allowed us to confirm the necessity result mentioned above. When all the constraints are imposed, a comparably small number of $761$ or $352$ out of the $2^{19}$ sign choices remain possible when ``$+$'' or ``$-$'' is fixed in \eqref{eqsolutionpm}, respectively. It then turns out that, by choosing an admissible sign configuration, all the numbers in the lists \eqref{eqelondiag} and \eqref{eqminors} can be made positive only if one of the quadruples listed in Table \ref{tabinvariants} consists solely of positive numbers. This is in full agreement with, and provides a rigorous, although not very insightful proof of the results reported in Section 5 of \cite{HHMS2011}. Actually, it turns out that there are precisely $12$ admissible sign configurations that correspond to a positive $\rho$ for some choice of the sign $\pm$ in \eqref{eqsolutionpm} and each of the quadruples in Table \ref{tabinvariants} is positive precisely for one of them. A complete list of the selected sign choices and the corresponding permutations is given in Appendix~\ref{appsignchoices}. Interestingly, $10$ of them correspond to choosing the plus sign in \eqref{eqsolutionpm}, while only $2$ to the minus sign. This is rather an uneven partitioning of the total of $12$ configurations, which is somewhat puzzling. 

To summarize, the computer-aided proof we carried out allows us to state the following.
\begin{proposition}\label{propdeterminationbygUPB}
A necessary and sufficient criterion for a generalized Unextendible Product Basis $\left\{\phi_i\otimes\psi_i\right\}_{i=1}^5\subset\mathbbm{C}^3\otimes\mathbbm{C}^3$ to belong to the kernel of a rank $4$ PPT state $\rho$ is that there exists a permutation of the vectors $\phi_i\otimes\psi_i$ that it yields all the values of the invariants $s_1$, $s_2$, $s_3$ and $s_4$, defined as in equations \eqref{invariants1}, \eqref{invariants2}, \eqref{invariant3} and \eqref{invariant4}, positive.  When checking positivity of $s_i$, it is possible to consider only the $12$ permutations, listed in Appendix \ref{apppermutations}, and the corresponding expressions for the invariants, given in Table \ref{tabinvariants}. 
\begin{proof}
First of all, let us note that a separable state $\rho$ cannot have a gUPB in its kernel, since it must have a product state in its range. Thus in the following we may always assume that $\rho$ is entangled.
Let us prove sufficiency first. If the invariants are positive for the permuted vectors $\phi'_i\otimes\psi'_i:=\phi_{\sigma\left(i\right)}\otimes\psi_{\sigma\left(i\right)}$, we know from Proposition~\ref{propinvariantspositiveortho2} that there exists a $\SLtt$ transformation $A\otimes B$ such that the transformed vectors $\left(A\otimes B\right)\phi'_{i}\otimes\psi'_{i}=\left(A\otimes B\right)\phi_{\sigma\left(i\right)}\otimes\psi_{\sigma\left(i\right)}$ are elements of an orthogonal UPB $\left\{v_i\otimes w_i\right\}_{i=1}^5$. With no loss of generality, we may assume that the vectors $v_i\otimes w_i$ are normalized to unity. In such case the projection
\begin{equation}\label{eqprojector}
\rho':=\mathbbm{1}-\sum_{i=1}^5\proj{v_i\otimes w_i}{v_i\otimes w_i}
\end{equation}
has all the vectors $v_i\otimes w_i$ in its kernel and it is a PPT entangled state \cite{Bennett99}. The locally transformed PPT state $\rho=\left(A\otimes B\right)^{\ast}\rho'\left(A\otimes B\right)$ has all the vectors $\phi_i\otimes\psi_i$ in its kernel.

In order to prove necessity, note that from the discussion above we know that positivity of $s_1,\ldots,s_4$, possibly afer a permutation, is a necessary condition for a PPT entangled state $\rho'$ with vectors $\phi_i\otimes\psi_i$ in its kernel to exist, provided that the vectors are as in equation \eqref{formtwoindeptriples4}. But any gUPB $\left\{\phi_i\otimes\psi_i\right\}_{i=1}^5$ can be brought to the form \eqref{formtwoindeptriples4} by a local transform, say $C\otimes D$. If we assume that a PPT state $\rho$ has $\left\{\phi_i\otimes\psi_i\right\}_{i=1}^5$ in its kernel, then the locally transformed $\rho'':=\left(C^{-1}\otimes D^{-1}\right)^{\ast}\rho\left(C^{-1}\otimes D^{-1}\right)$ has $\left(C\otimes D\right)\phi_i\otimes\psi_i$ in its kernel. But $\left(C\otimes D\right)\phi_i\otimes\psi_i$ are of the form \eqref{formtwoindeptriples4}. From the above discussion, $\rho''$ is PPT if and only if the invariants $s_1,\ldots,s_4$ are positive, possibly after we permute the vectors $\left(C\otimes D\right)\phi_i\otimes\psi_i$. But $C\otimes D$ does not change the value of the invariants, and thus $\phi_i\otimes\psi_i$, permuted in the same way as the $\left(C\otimes D\right)\phi_i\otimes\psi_i$, must also have all of them positive.
\end{proof}
\end{proposition}
Let us also state the following result, which should be expected from the discussion above.
\begin{proposition}\label{propuniqueness}
Let $\left\{\phi_i\otimes\psi_i\right\}_{i=1}^5\subset\mathbbm{C}^3\otimes\mathbbm{C}^3$ be a gUPB that yields, after a suitable permutation of the product vectors, positive values of all the invariants $s_1,\ldots,s_4$. The PPT state $\rho$ with $\left\{\phi_i\otimes\psi_i\right\}_{i=1}^5$ in its kernel is uniquely determined, up to scaling by a constant positive factor.
\begin{proof}
We already know that the assertion of the proposition holds for gUPBs of the form \eqref{formtwoindeptriples4}. We also know that any gUPB $\left\{\phi_i\otimes\psi_i\right\}_{i=1}^5$ can be locally transformed so that it looks like in \eqref{formtwoindeptriples4}. Let us denote the transformation which does it by $C\otimes D$. There cannot exist two PPT states $\rho_1$ and $\rho_2$ with $\left\{\phi_i\otimes\psi_i\right\}_{i=1}^5$ in their kernels, because in such case the PPT states $\left(C^{-1}\otimes D^{-1}\right)^{\ast}\rho_1\left(C^{-1}\otimes D^{-1}\right)$ and $\left(C^{-1}\otimes D^{-1}\right)^{\ast}\rho_2\left(C^{-1}\otimes D^{-1}\right)$ would both have the same gUPB of the form \eqref{formtwoindeptriples4} in their kernel, which we know is not possible.
\end{proof}
\end{proposition}
\subsection{The main result}\label{secmainresult}
Using the knowledge from the previous sections, we can now easily prove our main result.
\begin{theorem}\label{maintheorem}
Positive-partial-transpose states of rank $4$ in $3\times 3$ systems are either separable or they are of the form
\begin{equation}\label{eqlocaltransform}
\rho=\left(A\otimes B\right)^{\ast}\left(\mathbbm{1}-\sum_{i=1}^5\proj{v_i\otimes w_i}{v_i\otimes w_i}\right)\left(A\otimes B\right)
\end{equation}
with $A,B\in\SLt$ and $\left\{v_i\otimes w_i\right\}_{i=1}^5$ an orthonormal Unextendible Product Basis. In the latter case, they are entangled, and extreme in the set of PPT states. The rank of the partial transpose of the state is  $4$ in case of nonseparable states. 
\begin{proof}
In case of separable states, there is nothing to prove.
Let $\rho$ be a non-separable PPT state of rank $4$ in a $3\times 3$ system. We know from Proposition \ref{propgUPB} that there is a generalized UPB, say $\left\{\phi_i\otimes\psi_i\right\}_{i=1}^5$, in the kernel of $\rho$. From Proposition \ref{propdeterminationbygUPB} we know that the corresponding values of the invariants $s_1,\ldots,s_4$ must be all positive after we suitably permute the vectors $\phi_i\otimes\psi_i$. Next, Proposition \ref{propinvariantspositiveortho} tells us that there exists a $\SLtt$ transformation $A\otimes B$ that brings $\left\{\phi_i\otimes\psi_i\right\}_{i=1}^5$ to an orthogonal UPB $\left\{v_i\otimes w_i\right\}_{i=1}^5$. With no loss of generality, we may assume that the vectors $v_i\otimes w_i$ are normalized. From Proposition \ref{propuniqueness} we know that there exists, up to scaling, exactly one PPT state which has $\left\{v_i\otimes w_i\right\}_{i=1}^5$ in its kernel. It must be $\mathbbm{1}-\sum_{i=1}^5\proj{v_i\otimes w_i}{v_i\otimes w_i}$. The state given by the formula \eqref{eqlocaltransform} clearly is PPT, and it has all the vectors $\phi_i\otimes\psi_i$ in its kernel. By using Proposition~\ref{propuniqueness} again, we see that it must be equal to the $\rho$ we started with. The fact that the rank of the partial transpose is $4$ for non-separable states, is simply the assertion of Proposition \ref{propequalranks}.
\end{proof}
\end{theorem}  
In this way, we have obtained a full characterization of bound entangled states of minimal rank. Let us also mention a special property they have, which can be loosely described as saying that it is not enough for an entanglement witness to be indecomposable in order to detect them.
\begin{remark}\label{remarkatomicity}
According to \cite[Lemma 3]{SBL2001}, all PPT states of rank $4$ in $3\times 3$ systems can be written as a sum of four projections onto vectors of Schmidt rank $2$. By Theorem \ref{maintheorem}, or Proposition \ref{propequalranks}, their partial transposes are also of rank $4$ and thus can be decomposed in an analogous way. Using the notation of \cite{SSZ2009}, we can write that all such PPT states are elements of the cone $\mathcal{S}_{2,2}$. The dual cone $\mathcal{S}_{2,2}^{\circ}=\mathcal{D}_{2,2}$ consists of Jamiołkowski-Choi transforms of convex sums of $2$-positive and $2$-co-positive maps. Consequently, any entanglement witness that detects a PPT state of rank $4$ in a $3\times 3$ system is {\bf atomic} \cite{Ha98}. This applies in particular to the witness discussed in Example 1 of \cite{Terhal2001}.
\end{remark}

\subsection{Another use of product vectors} From the long argument above we have certainly learned that product vectors in the kernel play an important role for PPT states. It seems desirable to give an additional example of how they can be used to explain properties of PPT states reported in \cite{LS2010n}. We have the following
\begin{proposition}\label{propprodvecinrange}
A PPT state $\rho$ acting on $\mathbbm{C}^2\otimes\mathbbm{C}^n$ must have a product vector in its range.
\begin{proof}
Assume that there are no product vectors in $\range{\rho}$. According to Lemma \ref{lemmaAugusiak}, which originates from \cite[Lemmas 1 \& 2]{ATL2010}, the kernel of $\rho$, which is $\range{\rho}^{\bot}$, contains a set of product states $\left\{\phi_i\otimes\psi_i\right\}_{i=1}^6$ such that their partial conjugates $\phi_i^{\ast}\otimes\psi_i$ span $\mathbbm{C}^2\otimes\mathbbm{C}^n$. But this is impossible in our case, because according to Lemma \ref{lemmaconj}, all the vectors $\phi_i^{\ast}\otimes\psi_i$ must be in $\kernel{\rho^{T_1}}$. If they spanned $\mathbbm{C}^2\otimes\mathbbm{C}^n$, we would necessarily have $\rho^{T_1}=0$ as a consequence of positivity of the partial transpose. Thus we see that $\range{\rho}$ must contain a product vector, or in other words, it is not a CES.
\end{proof}
\end{proposition}
One can check in Table I of \cite{LS2010n} that all PPT states numerically studied by the authors in the $2\times 4$ and $2\times 5$ cases actually have a product vector in their range. The above proposition shows this is a general phenomenon. On the contrary, Tables~II -- IV of \cite{LS2010n} reveal PPT states, supportes on $3\times n$ and $4\times 4$ subspaces, that do not admit a product vector in their range.

\section{Conclusion}
It seems profitable to sometimes drop the orthogonality condition for Unextendible Product Bases, just as it is useful to extend the group of allowed transformations for PPT states from local special unitary to local special linear transformation \cite{LMO2006,LS2010}. By using this approach, with help of Bezout's Theorem from algebraic geometry, we have obtained a characterization of bound entangled PPT states of minimal rank, discovered earlier in numerical searches by the authors of \cite{LS2010}. The states are all $\SLtt$ transforms of projections onto  orthogonal complements of  orthogonal Unextendible Product Bases. In this sense, they are a ``deformed'' version of the bound entangled states introduced in \cite{Bennett99}. Certain methods we used apply to more general cases, and we hope they can help to explain even more peculiarities of the tables in \cite{LS2010n}. We already did explain a few of them, including the absence of non-separable PPT states $\rho$ of rank $4$ with the rank of $\rho^{T_1}$ different from $4$ and the fact that $2\times n$ PPT states must always have a product vector in their range. The use of product vectors in the kernel and the range of a PPT state is a common trait to most of the work reported above. Note that the paper \cite{AGKL2011} by other authors discusses extreme PPT states in the $2\times 4$ case, which can be regarded as another step towards an improved understanding of the results of \cite{LS2010n}.

One of our main aims  was to show that algebraic geometry can be useful in solving very concrete problems in quantum information theory. We could probably say that we went somewhat further than the papers \cite{Partha04,Cubitt07,Walgate08}, but in a similar spirit. Apart from solving the main example in Section \ref{sec2}, we showed, for instance, that some of the findings of \cite{ATL2010} can be found in a disguised form in a textbook like \cite{Harris}. We also used the opportunity to collect basic results about general Unextendible Product Bases, which seem to be missing from present literature. The gUPB framework arises in a natural way when local $\SL$ transformations are used instead of local unitary ones, and we could see this in Section \ref{sec2}.

\section*{Acknowledgement}
Interaction with Jon Magne Leinaas, Jan Myrheim, Per {\O}yvind Sollid and Andreas Hauge during my stay in Oslo and later on is gratefully acknowledged. I very much appreciate comments by Ingemar Bengtsson, Erling St{\o}rmer and Karol Życzkowski, which lead to a number of improvements in this paper. 

I am greatly indebted to the Mathematics Department of Oslo University and the Physics Department of Stockholm University for financially supporting my visity to Oslo. This work was also supported by the International PhD Projects Programme of the Foundation for Polish Science within the European Regional Development Fund of the European Union, agreement no. MPD/2009/6.

\bibliographystyle{unsrt}
\bibliography{non_ortho_UPBs}{}

\appendix
\section{Inequivalent permutations}\label{apppermutations}
This appendix gives a list of representatives of the $12$ equivalence classes of the symmertic group $S_5$ under left multiplication by the regular pentagram group, generated by the cycle $\left(1\,2\,3\,4\,5\right)$ and the axial symmetry $\left(\begin{array}{ccccc}1&2&3&4&5\\5&4&3&2&1\end{array}\right)$. We refer to them in Remark \ref{remarkpermutation} and in the discussion in Section \ref{secdetermination}. The list is in full compliance  with \cite{HHMS2011}, but we provide it here for completness.
\begin{eqnarray}
\sigma_1:\left(\begin{array}{ccccc}1&2&3&4&5\\1&2&3&4&5\end{array}\right)&\sigma_2:\left(\begin{array}{ccccc}1&2&3&4&5\\1&3&2&4&5\end{array}\right)&\sigma_3:\left(\begin{array}{ccccc}1&2&3&4&5\\2&1&3&4&5\end{array}\right)\nonumber\\
\sigma_4:\left(\begin{array}{ccccc}1&2&3&4&5\\2&3&1&4&5\end{array}\right)&\sigma_5:\left(\begin{array}{ccccc}1&2&3&4&5\\3&1&2&4&5\end{array}\right)&\sigma_6:\left(\begin{array}{ccccc}1&2&3&4&5\\3&2&1&4&5\end{array}\right)\nonumber\\
\sigma_7:\left(\begin{array}{ccccc}1&2&3&4&5\\1&2&4&3&5\end{array}\right)&\sigma_8:\left(\begin{array}{ccccc}1&2&3&4&5\\1&4&2&3&5\end{array}\right)&\sigma_9:\left(\begin{array}{ccccc}1&2&3&4&5\\2&1&4&3&5\end{array}\right)\nonumber\\
\sigma_{10}:\left(\begin{array}{ccccc}1&2&3&4&5\\2&4&1&3&5\end{array}\right)&\hskip -4 pt\sigma_{11}:\left(\begin{array}{ccccc}1&2&3&4&5\\1&3&4&2&5\end{array}\right)&\hskip -4 pt\sigma_{12}:\left(\begin{array}{ccccc}1&2&3&4&5\\1&4&3&2&5\end{array}\right)\nonumber
\end{eqnarray}


\section{Sign choices that yield a positive $\rho$}\label{appsignchoices}
In Table \ref{tabchoices}, we provide a complete list of twelve admissible sign choices for the numbers listed in equations \eqref{numerkilista1} and \eqref{numerkilista2} that yield a positive matrix \eqref{eqsolutionpm}. It turns out that for each sign configuration in Table \ref{tabchoices}, there is exactly one permutation $\sigma_i$ among those listed in Appendix \ref{apppermutations} that yields a positive value of all the invariants $s_1,\ldots,s_4$ defined by formulas \eqref{invariants1}, \eqref{invariants2}, \eqref{invariant3} and \eqref{invariant4}  (for more details, cf. Section \ref{secdetermination}). It should be mentioned that the rows in Table \ref{tabchoices} corresponding to $\sigma_9$ and $\sigma_{10}$ need a minus sign in \eqref{eqsolutionpm}, while all the other ones correspond to choosing a plus sign. To explain the notation we used in the header of Table \ref{tabchoices}, it is probably sufficient to say that the symbols $pp$, $qq$, $rr$ and $ss$ denote $p-1$, $q-1$, $r-1$ and $s-1$, respectively, while $pq$, $rs$, $pr$, $rq$, $qs$, $qrp$, $qrs$, $psq$, $rps$ and $qrps$ stand for $p-q$, $r-s$, $p-r$, $r-q$, $qr-p$, $qr-s$, $ps-q$, $r-ps$ and $qr-ps$, respectively.
\begin{table}[ht]
\scalebox{0.85}{
\begin{tabular}{|>{$}c<{$}|c|c|c|c|c|c|c|c|c|c|c|c|c|c|c|c|c|c|c|}
\cline{2-20}
\multicolumn{1}{c|}{}&p&q&r&s&pp&qq&rr&ss&pq&rs&pr&ps&rq&qs&qrp&qrs&psq&rps&qrps\\
\hline
\sigma_1&--&+&+&+&--&+&--&--&--&+&--&--&--&+&+&+&--&+&+\\
\sigma_2&+&--&+&+&+&--&--&--&+&--&+&+&+&--&--&--&+&--&--\\
\sigma_3&--&--&+&+&--&--&+&--&+&+&--&--&+&--&--&--&+&+&--\\
\sigma_4&+&--&+&+&--&--&+&+&+&+&--&--&+&--&--&--&+&+&--\\
\sigma_5&--&--&+&+&--&--&--&+&--&--&--&--&+&--&+&--&--&+&+\\
\sigma_6&--&+&+&+&--&--&+&+&--&--&--&--&+&--&+&--&--&+&+\\
\sigma_7&+&+&+&--&+&+&--&--&+&+&+&+&--&+&--&+&--&+&+\\
\sigma_8&+&+&--&--&+&--&--&--&+&+&+&+&--&+&--&+&--&+&+\\
\sigma_9&+&+&+&--&--&--&+&--&--&+&--&+&+&+&+&+&--&+&+\\
\sigma_{10}&+&+&--&+&--&--&--&+&+&--&+&--&--&--&--&--&+&--&--\\
\sigma_{11}&+&+&--&+&+&+&--&--&--&--&+&+&+&+&--&--&--&--&--\\
\sigma_{12}&+&+&--&--&--&+&--&--&--&--&+&+&--&+&--&--&--&--&--\\
\hline
\end{tabular}
}
\vskip 2 pt
\caption{Sign choices that yield a positive $\rho$ and obey all the constraints of Appendix \ref{appconstraints}.}
\label{tabchoices}
\end{table}

\section{Sign constraints}\label{appconstraints} 
In this appendix, we give a list of constraints on the numbers listed in equations \eqref{numerkilista1} and \eqref{numerkilista2}. They are contained in Tables \ref{tabsigns3} and \ref{tabsigns2}. Some of the constraints are simple relations like $p-1>0\Rightarrow p>0$, but there are also more sophisticated ones. For example,
\begin{equation}\label{sophistic}
\left(r<0\land q-1<0\land r-ps>0\right)\Rightarrow qr-ps>0.
\end{equation}
Alternatively, the above formula can be written as
\begin{equation}\label{sophistic2}
\neg\left(r<0\land q-1<0\land r-ps>0\land qr-ps<0\right).
\end{equation}
This is also the convention we adopt in the following tables. For example, formula \eqref{sophistic2} corresponds to the following row in Table \ref{tabsigns3}.
\begin{equation}\nonumber
\scalebox{0.9}{
\begin{tabular}{|c|c|c|c|c|c|c|c|c|c|c|c|c|c|c|c|c|c|c|}
\hline
p&q&r&s&pp&qq&rr&ss&pq&rs&pr&ps&rq&qs&qrp&qrs&psq&rps&qrps\\
\hline
&&--&&&--&&&&&&&&&&&&+&--\\
\hline
\end{tabular}
}
\end{equation}
An explanation of the notation we used in the headers of the tables can be found in Appendix \ref{appsignchoices}.

\begin{table}[b]
\scalebox{0.9}{
\begin{tabular}{|c|c|c|c|c|c|c|c|c|c|c|c|c|c|c|c|c|c|c|}
\hline
p&q&r&s&pp&qq&rr&ss&pq&rs&pr&ps&rq&qs&qrp&qrs&psq&rps&qrps\\
\hline
&&+&&&+&&&&+&&&&&&--&&&\\
&+&&&&&+&&&&&&&+&&--&&&\\
&&+&&&--&&&&--&&&&&&+&&&\\
&+&&&&&--&&&&&&&--&&+&&&\\
&&&+&+&&&&&&&&&--&&&--&&\\
+&&&&&&&+&+&&&&&&&&--&&\\
&&&+&--&&&&&&&&&+&&&+&&\\
+&&&&&&&--&--&&&&&&&&+&&\\
&+&&&&&+&&--&&&&&&--&&&&\\
&&+&&&+&&&&&--&&&&--&&&&\\
&+&&&&&--&&+&&&&&&+&&&&\\
&&+&&&--&&&&&+&&&&+&&&&\\
+&&&&&&&+&&&+&&&&&&&+&\\
&&&+&+&&&&&--&&&&&&&&+&\\
+&&&&&&&--&&&--&&&&&&&--&\\
&&&+&--&&&&&+&&&&&&&&--&\\
&&&+&--&&&&&&&&&&&+&&&--\\
&&&+&+&&&&&&&&&&&--&&&+\\
&+&&&&&+&&&&&&&&&&--&&--\\
&+&&&&&--&&&&&&&&&&+&&+\\
+&&&&&&&--&&&&&&&+&&&&--\\
+&&&&&&&+&&&&&&&--&&&&+\\
&&+&&&+&&&&&&&&&&&&+&--\\
&&+&&&--&&&&&&&&&&&&--&+\\
&&--&&&+&&&&--&&&&&&+&&&\\
&&--&&&--&&&&+&&&&&&--&&&\\
&--&&&&&--&&&&&&&+&&--&&&\\
&--&&&&&+&&&&&&&--&&+&&&\\
&&&--&--&&&&&&&&&--&&&--&&\\
--&&&&&&&--&+&&&&&&&&--&&\\
&&&--&+&&&&&&&&&+&&&+&&\\
--&&&&&&&+&--&&&&&&&&+&&\\
&--&&&&&--&&--&&&&&&--&&&&\\
&&--&&&--&&&&&--&&&&--&&&&\\
&--&&&&&+&&+&&&&&&+&&&&\\
&&--&&&+&&&&&+&&&&+&&&&\\
--&&&&&&&+&&&--&&&&&&&--&\\
&&&--&+&&&&&+&&&&&&&&--&\\
--&&&&&&&--&&&+&&&&&&&+&\\
&&&--&--&&&&&--&&&&&&&&+&\\
&&&--&+&&&&&&&&&&&+&&&--\\
&&&--&--&&&&&&&&&&&--&&&+\\
&--&&&&&--&&&&&&&&&&--&&--\\
&--&&&&&+&&&&&&&&&&+&&+\\
--&&&&&&&+&&&&&&&+&&&&--\\
--&&&&&&&--&&&&&&&--&&&&+\\
&&--&&&--&&&&&&&&&&&&+&--\\
&&--&&&+&&&&&&&&&&&&--&+\\
\hline
\end{tabular}
}
\vskip 2 pt
\caption{Non-admissible sign choices. Part I.}
\label{tabsigns3}
\end{table}

\begin{table}[b]
\scalebox{0.9}{
\begin{tabular}{|c|c|c|c|c|c|c|c|c|c|c|c|c|c|c|c|c|c|c|}
\hline
p&q&r&s&pp&qq&rr&ss&pq&rs&pr&ps&rq&qs&qrp&qrs&psq&rps&qrps\\
\hline
--&&&&+&&&&&&&&&&&&&&\\
&--&&&&+&&&&&&&&&&&&&\\
&&--&&&&+&&&&&&&&&&&&\\
&&&--&&&&+&&&&&&&&&&&\\
--&+&&&&&&&+&&&&&&&&&&\\
+&--&&&&&&&--&&&&&&&&&&\\
&&+&--&&&&&&--&&&&&&&&&\\
&&--&+&&&&&&+&&&&&&&&&\\
&&&&--&+&&&+&&&&&&&&&&\\
&&&&+&--&&&--&&&&&&&&&&\\
&&&&&&--&+&&+&&&&&&&&&\\
&&&&&&+&--&&--&&&&&&&&&\\
%
&&&&&--&&+&&&&&&+&&&&&\\
&--&&+&&&&&&&&&&+&&&&&\\
&+&&--&&&&&&&&&&--&&&&&\\
&&&&&+&&--&&&&&&--&&&&&\\
--&&+&&&&&&&&+&&&&&&&&\\
&&&&--&&+&&&&+&&&&&&&&\\
+&&--&&&&&&&&--&&&&&&&&\\
&&&&+&&--&&&&--&&&&&&&&\\
&+&--&&&&&&&&&&+&&&&&&\\
&&&&&+&--&&&&&&+&&&&&&\\
&--&+&&&&&&&&&&--&&&&&&\\
&&&&&--&+&&&&&&--&&&&&&\\
--&&&+&&&&&&&&+&&&&&&&\\
&&&&--&&&+&&&&+&&&&&&&\\
+&&&--&&&&&&&&--&&&&&&&\\
&&&&+&&&--&&&&--&&&&&&&\\
\hline
\end{tabular}
}
\vskip 2 pt
\caption{Non-admissible sign choices. Part II.}
\label{tabsigns2}
\end{table}

\end{document}